\documentclass[a4paper]{article}
\usepackage{setspace}
\usepackage[margin=1in]{geometry}
\usepackage{amsmath, amssymb, graphicx, tabularx, float, enumerate, amsthm}
\usepackage[sectionbib]{natbib}
\usepackage[para]{threeparttable}
\newtheorem{theorem}{Theorem}
\newtheorem{lemma}{Lemma}
\usepackage[title]{appendix}
\allowdisplaybreaks[2]

\providecommand{\keywords}[1]
{
  \small	
  \textbf{\textit{Keywords:}} #1
}

\begin{document}

\title{A global kernel estimator for partially linear varying coefficient additive hazards models}

\author{Hoi Min Ng and Kin Yau Wong\thanks{Corresponding author; email: kin-yau.wong@polyu.edu.hk}\\[4pt]Department of Applied Mathematics, The Hong Kong Polytechnic University}

\date{}

\maketitle

\begin{abstract}
In biomedical studies, we are often interested in the association between different types of covariates and the times to disease events. Because the relationship between the covariates and event times is often complex, standard survival models that assume a linear covariate effect are inadequate. A flexible class of models for capturing complex interaction effects among types of covariates is the varying coefficient models, where the effects of a type of covariates can be modified by another type of covariates. In this paper, we study kernel-based estimation methods for varying coefficient additive hazards models. Unlike many existing kernel-based methods that use a local neighborhood of subjects for the estimation of the varying coefficient function, we propose a novel global approach that is generally more efficient. We establish theoretical properties of the proposed estimators and demonstrate their superior performance compared with existing local methods through large-scale simulation studies. To illustrate the proposed method, we provide an application to a motivating cancer genomic study.
\end{abstract}


\section{Introduction}

In biomedical studies, we are often interested in the association between covariates and the time to some disease event, such as disease recurrence or death. A popular approach is to model the effect of the covariates on the event time through the hazard function, which represents the instantaneous risk of the event at a time point given that the event has not yet occurred. Such models capture the dynamics of the event occurrence process. Among such models, the most popular choice is the proportional hazards model \citep{cox1972regression}, which assumes that the covariates act multiplicatively on the hazard function. The popularity of the proportional hazards model is mainly due to the intuitive interpretation of the regression parameters and the availability of the partial likelihood, which enables a simple procedure for estimation and inference of the regression parameters without parametric assumptions on the baseline hazard function. Another model that directly captures the effects of covariates on the hazard is the additive hazards model \citep{aalen1989linear,lin1994semiparametric}, which assumes that the covariates act additively to the hazard function. In particular, the conditional hazard function of an event time given a possibly time-varying covariate $\boldsymbol{X}(\cdot)$ takes the form
\[
    \lambda(t\mid \boldsymbol{X})=\lambda(t)+\boldsymbol{\beta}^{\mathrm{T}}\boldsymbol{X}(t),
\]
where $\lambda(\cdot)$ is an unspecified positive function and $\boldsymbol{\beta}$ represents the covariate effect. Since the proportional hazards and additive hazards models assume different structures of association between the covariates and event time, both models would be of practical interest, especially in the data exploration stage. While the choice between the models should be made based primarily on the scientific question of interest and the empirical fit of data, one advantage of the additive hazards model is the availability of closed-form estimators of the unknown parameters.

In some applications, the outcome could be influenced by the interplay among various covariates. For example, disease outcomes are often influenced by complex biological processes that involve various genomic features. Also, \cite{hunter2005gene} suggested that the effect of a genotype on the risk of disease development may differ based on environmental factors. In addition, \cite{landi2008gene} demonstrated that the effects of some gene expressions on the risk of lung cancer mortality vary with tobacco consumption. To account for such interaction effects among covariates, we can formulate $\boldsymbol{\beta}$ as a function of another covariate $\boldsymbol{W}$. This varying coefficient structure has been widely applied to various models, including (generalized) linear models \citep{hastie1993varying,fan1999statistical,fan2006local}, proportional hazards models \citep{tian2005cox}, and additive hazards models \citep{aalen1989linear,mckeague1994partly,yin2008partially,li2007local,qu2018identification}.

Varying coefficient additive hazards models generally take the form
\begin{align}
\label{eq:vc model with t}
    \lambda(t\mid \boldsymbol{W},\boldsymbol{X})=\lambda(t)+\boldsymbol{\beta}(\boldsymbol{W}(t))^{\mathrm{T}}\boldsymbol{X}(t),
\end{align}
where $\boldsymbol{\beta}(\boldsymbol{w})$ is a vector of coefficient functions that represents the effect of $\boldsymbol{X}(t)$ at $\boldsymbol{W}(t)=\boldsymbol{w}$. In this model, the effect of $\boldsymbol{X}(t)$ is modified by $\boldsymbol{W}(t)$, representing a possibly nonlinear interaction between $\boldsymbol{W}(t)$ and each component of $\boldsymbol{X}(t)$. Model~\eqref{eq:vc model with t} includes various models as special cases. For example, when $W(t)$ is a univariate variable that represents time, that is, $W(t)=t$, the model reduces to the additive hazards model with time-varying coefficients \citep{aalen1989linear}. This model allows the covariate effects to vary over time, capturing time-dependent relationships between the covariates and the hazard rate. When the coefficient $\boldsymbol{\beta}$ is constant in model~\eqref{eq:vc model with t}, the model simplifies to the additive hazards model with constant coefficients \citep{lin1994semiparametric}. Partially linear models can also be considered within the framework of model~\eqref{eq:vc model with t} by setting some components of $\boldsymbol{\beta}$ to be constant \citep{mckeague1994partly,yin2008partially}.

In this paper, we focus on kernel smoothing techniques for the estimation of varying coefficient additive hazard models. The basic idea underlying kernel methods is to estimate a function at any given value based on subjects associated with a local neighborhood of the value. Kernel-based methods have been extensively applied to the estimation of varying coefficient survival models, such as the proportional hazards models \citep{tian2005cox,fan2006local} and the additive hazards models \citep{li2007local,yin2008partially,qu2018identification}. In these methods, the proposed estimators are obtained by taking existing estimators (for non-varying coefficient models) and including the kernel function as a weight to each observation. Because these estimators for the varying coefficient function at any point are based on a small local neighborhood of subjects, we generally refer to them as ``local'' kernel estimation methods. While local methods are easy to implement and can be straight-forwardly applied to various semiparametric models, they may suffer efficiency loss because information contained in subjects outside the neighborhood is discarded. To reduce the efficiency loss, \cite{chen2012efficient} proposed global kernel estimation methods for the varying coefficient proportional hazards models that includes all subjects in the estimation of the varying coefficients. While their estimators improve efficiency, their methods could be computationally demanding due to the iterative nature of the algorithm.

We propose a global kernel estimator for the (partially linear) varying coefficient additive hazards model. Instead of separately estimating the varying coefficient function at any given covariate value using a local neighborhood of subjects, the proposed method estimates the varying coefficient function over a range of covariate values simultaneously and thus involves all subjects. The local method suffers efficiency loss because, effectively, only a small neighborhood of subjects are used to simultaneously estimate the varying coefficient function at any given value and the baseline hazard function. However, the baseline hazard function is shared among all subjects and could be more efficiently estimated by using all subjects. The proposed method takes into account this sharing of a common baseline hazard function to improve efficiency. The proposed estimator is the solution to a system of linear equations and has a closed-form solution, and thus it is computationally more efficient than existing global kernel estimators.

\section{Model and estimation}

\subsection{Additive hazards model with varying covariate effects}

We first consider an additive hazards model with only varying covariate effects. Let $\boldsymbol{W}\equiv(W_1,\ldots,W_q)^\mathrm{T}$ and $\boldsymbol{X}\equiv(X_1,\ldots,X_p)^\mathrm{T}$ be two sets of covariates and $\widetilde{T}$ be the event time. We assume the following hazard function for $\widetilde{T}$:
\begin{align}
\label{eq:vc model}
    \lambda(t\mid \boldsymbol{W},\boldsymbol{X})=\lambda(t)+\boldsymbol{\beta}(\boldsymbol{W})^{\mathrm{T}}\boldsymbol{X}.
\end{align}
Note that in contrast with model~\eqref{eq:vc model with t}, we focus on time-independent covariates. We allow $\widetilde{T}$ to be right-censored and let $C$ be the censoring time. Assume that $C$ and $\widetilde{T}$ are conditionally independent given $(\boldsymbol{W},\boldsymbol{X})$. Let $T\equiv\min(\widetilde{T},C)$ be the observed time and $\Delta\equiv I(\widetilde{T}\leq C)$ be the event indicator. For a sample of size $n$, we observe $(T_i,\Delta_i,\boldsymbol{W}_i,\boldsymbol{X}_i)$ for $i=1,\ldots,n$. Define $N_i(t)\equiv\Delta_i I(T_{i}\leq t)$ and $Y_{i}(t)\equiv I(T_{i}\geq t)$ as the observed event process and at-risk process for the $i$th subject at time $t$, respectively. In model~\eqref{eq:vc model}, if $\boldsymbol{\beta}$ is constant, then we can estimate it using the estimator of \cite{lin1994semiparametric}:
\begin{align*}
    \bigg[\sum^{n}_{i=1}\int^{\tau}_{0} Y_{i}(t)\big\{\boldsymbol{X}_{i}-\overline{\boldsymbol{X}}(t)\big\}^{\otimes 2}\,\mathrm{d}t\bigg]^{-1}\sum^{n}_{i=1}\int^{\tau}_{0}\big\{\boldsymbol{X}_{i}-\overline{\boldsymbol{X}}(t)\big\}\,\mathrm{d}N_{i}(t),
\end{align*}
where $\tau$ is the maximum follow-up time, $\overline{\boldsymbol{X}}(t)\equiv\sum^{n}_{i=1}\boldsymbol{X}_{i}Y_{i}(t)/\sum^{n}_{i=1}Y_{i}(t)$ is the average covariate vector among subjects at risk at time $t$ with the convention that $0/0=0$, and for any vector $\boldsymbol{a}$, $\boldsymbol{a}^{\otimes 2}=\boldsymbol{a}\boldsymbol{a}^{\mathrm{T}}$.

To motivate the proposed estimation method for model~\eqref{eq:vc model}, we first consider the special case that $\boldsymbol{W}$ takes discrete values $\boldsymbol{w}_1,\ldots,\boldsymbol{w}_m$. In this case, we can define a vector of covariates $\boldsymbol{B}\equiv(I(\boldsymbol{W}=\boldsymbol{w}_1)\boldsymbol{X}^{\mathrm{T}},\ldots,I(\boldsymbol{W}=\boldsymbol{w}_m)\boldsymbol{X}^{\mathrm{T}})^{\mathrm{T}}$. Fitting the standard additive hazards model with covariate $\boldsymbol{B}$ is equivalent to fitting model~\eqref{eq:vc model}, and the Lin and Ying estimator can be applied to estimate $\boldsymbol{\beta}\equiv(\boldsymbol{\beta}(\boldsymbol{w}_1)^\mathrm{T},\ldots,\boldsymbol{\beta}(\boldsymbol{w}_m)^\mathrm{T})^\mathrm{T}$. The explicit formulation of the resulting estimator is provided in Appendix~\ref{sec:discrete}. For the case of a continuous $\boldsymbol{W}$, we mimic the estimator in the discrete case and apply the kernel smoothing technique. We define a grid $(\boldsymbol{w}_1,\ldots,\boldsymbol{w}_m)$ over the support of $\boldsymbol{W}$ and estimate the values of $\boldsymbol{\beta}$ over this grid simultaneously. In particular, we estimate $(\boldsymbol{\beta}(\boldsymbol{w}_1)^\mathrm{T},\ldots,\boldsymbol{\beta}(\boldsymbol{w}_m)^\mathrm{T})^\mathrm{T}$ by $\widehat{\boldsymbol{\beta}}\equiv\boldsymbol{V}^{-1}\boldsymbol{b}$, where $\boldsymbol{V}$ is a block matrix with $\boldsymbol{V}(\boldsymbol{w}_k,\boldsymbol{w}_\ell)$ as its $(k,\ell)$th block, $\boldsymbol{b}=(\boldsymbol{b}(\boldsymbol{w}_{1})^{\mathrm{T}},\ldots,\boldsymbol{b}(\boldsymbol{w}_m)^{\mathrm{T}})^{\mathrm{T}}$, and
\begin{align*}
    \boldsymbol{b}(\boldsymbol{w}_k)=&\frac{1}{nm}\sum_{i=1}^n\int_0^{\tau}\bigg\{K_H(\boldsymbol{W}_i-\boldsymbol{w}_k)\boldsymbol{X}_i-\sum_{j=1}^m K_H(\boldsymbol{W}_i-\boldsymbol{w}_j)\overline{\boldsymbol{X}}(t,\boldsymbol{w}_k)\bigg\}\,\mathrm{d}N_i(t),\\
    \boldsymbol{V}(\boldsymbol{w}_k,\boldsymbol{w}_\ell)=&\frac{I(\boldsymbol{w}_k=\boldsymbol{w}_\ell)}{nm}\sum_{i=1}^n\int_0^{\tau}K_H(\boldsymbol{W}_i-\boldsymbol{w}_k)Y_i(t)\boldsymbol{X}_i^{\otimes 2}\,\mathrm{d}t\\
    &-\frac{1}{nm}\sum_{i=1}^n\int_0^\tau\sum_{j=1}^m K_H(\boldsymbol{W}_i-\boldsymbol{w}_j)Y_i(t)\overline{\boldsymbol{X}}(t,\boldsymbol{w}_k)\overline{\boldsymbol{X}}(t,\boldsymbol{w}_\ell)^{\mathrm{T}}\,\mathrm{d}t,\\
    \overline{\boldsymbol{X}}(t,\boldsymbol{w}_k)=&\frac{\sum_{i=1}^n K_H(\boldsymbol{W}_i-\boldsymbol{w}_k)Y_i(t)\boldsymbol{X}_i}{\sum_{i=1}^n\sum_{j=1}^m K_H(\boldsymbol{W}_i-\boldsymbol{w}_j)Y_i(t)}.
\end{align*}
In the above formulation, $K_H(\boldsymbol{x})\equiv\vert\boldsymbol{H}\vert^{-1/2}K(\boldsymbol{H}^{-1/2}\boldsymbol{x})$ is a kernel function with bandwidth matrix $\boldsymbol{H}$. Essentially, we take the estimator for the discrete case (given in Appendix~\ref{sec:discrete}) and replace the summations over the subjects with $\boldsymbol{W}_i=\boldsymbol{w}_k$ by a weighted sum over all subjects, where the weight is defined based on the distance between individual $\boldsymbol{W}_i$'s and $\boldsymbol{w}_k$. For summations over all subjects in the estimator of the discrete case, one possible approach is to keep them unchanged in the continuous case. Nevertheless, we propose to multiply $m^{-1}\sum_{j=1}^m K_H(\boldsymbol{W}_i-\boldsymbol{w}_j)$ to each term in the summation. Although the two resulting estimators are asymptotically equivalent, in our experience, the estimator with $m^{-1}\sum_{j=1}^m K_H(\boldsymbol{W}_i-\boldsymbol{w}_j)$ is more robust to the choice of the grid.

From the proof of Theorem~\ref{thm:pointest} presented in Appendix~\ref{sec:proof1}, we can see that a consistent estimator of $\boldsymbol{\beta}(\cdot)$ can be obtained by solving an equation $\mathcal{V}\boldsymbol{\beta}=\mathcal{C}$, where $\mathcal{V}$ is a linear operator and $\mathcal{C}$ is a function that depends on the observed data. The proposed method effectively takes this equation and replaces the left-hand side by the product between a matrix and $\boldsymbol{\beta}$ evaluated at the grid points and replaces the right-hand side by a vector that consists of $\mathcal{C}$ evaluated at the grid points. Essentially,
we approximate the integral equations involved in $\mathcal{V}\boldsymbol{\beta}$ by summations over the grid points and solve for $\widehat{\boldsymbol{\beta}}$ at these grid points only. Increasing the number of grid points generally reduces the approximation error but increases the computational cost. To preserve computational efficiency, we choose a set of sparse grid points that adequately represents the distribution of $\boldsymbol{W}$. We suggest using a quantile-based grid, in which the grid points are positioned at the empirical quantiles of $\boldsymbol{W}$. Alternatively, an evenly spaced grid that uniformly covers the support of $\boldsymbol{W}$ can be adopted. For values of $\boldsymbol{\beta}$ outside the grid points, we can estimate them using linear interpolation based on  the values of $\widehat{\boldsymbol{\beta}}(\boldsymbol{w})$ at the nearest grid points.

When $m=1$, the proposed estimator reduces to a local estimator, which is similar in spirit to the estimator of \cite{yin2008partially}. In particular, $\boldsymbol{\beta}(\boldsymbol{w})$ is estimated by $\boldsymbol{V}(\boldsymbol{w})^{\mathrm{-1}}\boldsymbol{b}(\boldsymbol{w})$, where 
\begin{align*}
    \boldsymbol{b}(\boldsymbol{w})=&\frac{1}{n}\sum^{n}_{i=1}\int^{\tau}_{0}K_H(\boldsymbol{W}_{i}-\boldsymbol{w})\bigg\{\boldsymbol{X}_{i}-\frac{\sum^{n}_{i=1}K_H(\boldsymbol{W}_{i}-\boldsymbol{w})Y_{i}(t)\boldsymbol{X}_{i}}{\sum^{n}_{i=1}K_H(\boldsymbol{W}_{i}-\boldsymbol{w})Y_{i}(t)}\bigg\}\,\mathrm{d}N_{i}(t),\\
    \boldsymbol{V}(\boldsymbol{w})=&\frac{1}{n}\sum^{n}_{i=1}\int^{\tau}_{0}K_H(\boldsymbol{W}_{i}-\boldsymbol{w})Y_{i}(t)\bigg\{\boldsymbol{X}_{i}-\frac{\sum^{n}_{i=1}K_H(\boldsymbol{W}_{i}-\boldsymbol{w})Y_{i}(t)\boldsymbol{X}_{i}}{\sum^{n}_{i=1}K_H(\boldsymbol{W}_{i}-\boldsymbol{w})Y_{i}(t)}\bigg\}^{\otimes 2}\,\mathrm{d}t.
\end{align*}
This approach takes the Lin and Ying estimator and attaches a weight $K_H(\boldsymbol{W}_i-\boldsymbol{w})$ to the $i$th subject ($i=1,\ldots,n$); if $K_H(\boldsymbol{W}_i-\boldsymbol{w})$ is a constant for all $\boldsymbol{w}$, then the estimator reduces to the Lin and Ying estimator. The local kernel estimation method ``profiles out'' the baseline hazard function separately for each $\boldsymbol{w}$, thus implicitly assuming that the baseline hazard function varies with $\boldsymbol{W}$. The local method suffers loss of efficiency as it does not make use of the fact that the baseline hazard function is indeed shared. In the proposed method, $\boldsymbol{\beta}(\boldsymbol{w}_k)$'s are not estimated separately but are obtained simultaneously from a set of linear equations, because all subjects are jointly used to profile out the common baseline hazard function. Therefore, the global estimator tends to be more efficient than the local estimator.

In this paper, we consider a Gaussian kernel of $K(\boldsymbol{x})\propto\exp(\boldsymbol{x}^{\mathrm{T}}\boldsymbol{x}/2)$. A full bandwidth matrix $\boldsymbol{H}$ gives more flexibility but also quadratically increases the number of bandwidth parameters to be chosen, precisely $q(q+1)/2$. This complicates the bandwidth selection as $q$ grows. A common simplification is to consider a diagonal bandwidth matrix. Here, we adopt a diagonal bandwidth matrix of $\boldsymbol{H}=\mathrm{diag}(h_1^2,\ldots,h_q^2)$ and follow \cite{silverman1986density} to set $h_{j}=\widehat{\sigma}_{j}[4/\{n(q+2)\}]^{1/(q+4)}$, where $\widehat{\sigma}_{j}$ is the empirical standard deviation of $\boldsymbol{W}_{\cdot j}$ for $j=1,\ldots,q$. If computationally feasible, cross-validation can be applied to choose the bandwidth.

\subsection{Additive hazards model with mixed covariate effects}

In practice, certain covariate effects may be known to be constant. We can generally include a covariate vector $\boldsymbol{Z}$ that has a constant effect on the hazard rate and consider the following partially linear varying coefficient additive hazards model:
\begin{align*}
    \lambda(t\mid\boldsymbol{W},\boldsymbol{X},\boldsymbol{Z})=\lambda(t)+\boldsymbol{\beta}(\boldsymbol{W})^{\mathrm{T}}\boldsymbol{X}+\boldsymbol{\alpha}^{\mathrm{T}}\boldsymbol{Z},
\end{align*}
where $\boldsymbol{\alpha}$ is a vector of regression coefficients. We assume that the covariates $\boldsymbol{X}$ and $\boldsymbol{Z}$ do not overlap. In analogy with the kernel-weighted estimator motivated by the block-wise formulation for estimating the varying covariate effect, we propose to estimate $\boldsymbol{\beta}$ and $\boldsymbol{\alpha}$ simultaneously by
\begin{align}
\label{eq:estimator}
    \left(\begin{array}{c}
    \widehat{\boldsymbol{\beta}}\\
    \widehat{\boldsymbol{\alpha}}
    \end{array}\right)&=\left(\begin{array}{cc}
    \boldsymbol{V} & \boldsymbol{V}_{\beta\alpha}\\
    \boldsymbol{V}_{\beta\alpha}^{\mathrm{T}} & \boldsymbol{V}_{\alpha\alpha}
    \end{array}\right)^{-1}
    \left(\begin{array}{c}
    \boldsymbol{b}\\
    \boldsymbol{b}_\alpha
    \end{array}\right),
\end{align}
where $\boldsymbol{V}_{\beta\alpha}=(\boldsymbol{V}_{\beta\alpha}(\boldsymbol{w}_1)^{\mathrm{T}},\ldots,\boldsymbol{V}_{\beta\alpha}(\boldsymbol{w}_m)^{\mathrm{T}})^{\mathrm{T}}$
with
\begin{align*}
    \boldsymbol{b}_\alpha=&\frac{1}{nm}\sum^{n}_{i=1}\int^{\tau}_{0}\sum_{j=1}^m K_H(\boldsymbol{W}_i-\boldsymbol{w}_j)\big\{\boldsymbol{Z}_{i}-\overline{\boldsymbol{Z}}(t)\big\}\,\mathrm{d}N_{i}(t),\\
    \boldsymbol{V}_{\alpha\alpha}=&\frac{1}{nm}\sum^{n}_{i=1}\int^{\tau}_{0}\sum_{j=1}^m K_H(\boldsymbol{W}_i-\boldsymbol{w}_j)Y_{i}(t)\big\{\boldsymbol{Z}_{i}-\overline{\boldsymbol{Z}}(t)\big\}^{\otimes 2}\,\mathrm{d}t,\\
    \boldsymbol{V}_{\beta\alpha}(\boldsymbol{w}_k)=&\frac{1}{nm}\sum^{n}_{i=1}\int^{\tau}_{0}K_H(\boldsymbol{W}_{i}-\boldsymbol{w}_k)Y_{i}(t)\boldsymbol{X}_{i}\big\{\boldsymbol{Z}_{i}-\overline{\boldsymbol{Z}}(t)\big\}^{\mathrm{T}}\,\mathrm{d}t,\\
    \overline{\boldsymbol{Z}}(t)=&\frac{\sum_{j=1}^m\sum^{n}_{i=1}K_H(\boldsymbol{W}_i-\boldsymbol{w}_j)Y_{i}(t)\boldsymbol{Z}_{i}}{\sum_{j=1}^m\sum^{n}_{i=1}K_H(\boldsymbol{W}_i-\boldsymbol{w}_j)Y_{i}(t)}.
\end{align*}

Although $\boldsymbol{\alpha}$ does not vary with $\boldsymbol{W}$, the estimator $\widehat{\boldsymbol{\alpha}}$ depends on the chosen set of grid points for $\boldsymbol{W}$. To obtain a more robust estimator, we could update the estimator of $\boldsymbol{\alpha}$ after obtaining an estimator for $\boldsymbol{\beta}(\boldsymbol{W}_i)$ for each $i=1,\ldots,n$. The updated estimator is essentially the Lin and Ying estimator with $\boldsymbol{\beta}(\boldsymbol{W}_i)$ fixed at $\widehat{\boldsymbol{\beta}}(\boldsymbol{W}_i)$. In particular, we define the estimator as
\[
    \widetilde{\boldsymbol{\alpha}}\equiv\bigg[\sum_{i=1}^n\int_0^\tau Y_i(t)\big\{\boldsymbol{Z}_i-\widetilde{\boldsymbol{Z}}(t)\big\}^{\otimes 2}\,\mathrm{d}t\bigg]^{-1}\sum_{i=1}^n\int_0^\tau\big\{\boldsymbol{Z}_i-\widetilde{\boldsymbol{Z}}(t)\big\}\big\{\mathrm{d}N_i(t)-Y_i(t)\widehat{\boldsymbol{\beta}}(\boldsymbol{W}_i)^\mathrm{T}\boldsymbol{X}_i\,\mathrm{d}t\big\},
\]
where $\widetilde{\boldsymbol{Z}}(t)=\sum_{i=1}^n Y_i(t)\boldsymbol{Z}_i/\sum_{i=1}^n Y_i(t)$. This updated estimator $\widetilde{\boldsymbol{\alpha}}$ is more robust to the choice of grid. The cumulative baseline hazard function $\Lambda(t)\equiv\int_0^t \lambda(s)\,\mathrm{d}s$ can then be estimated by
\[
    \widehat{\Lambda}(t)=\int_0^t\frac{\sum_{i=1}^n\big\{\mathrm{d}N_i(s)-Y_i(s)\widehat{\boldsymbol{\beta}}(\boldsymbol{W}_i)^{\mathrm{T}}\boldsymbol{X}_i\,\mathrm{d}s-Y_i(s)\widetilde{\boldsymbol{\alpha}}^\mathrm{T}\boldsymbol{Z}_i\,\mathrm{d}s\big\}}{\sum_{i=1}^n Y_i(s)}.
\]
In the numerical studies, we used linear interpolation to approximate $\widehat{\boldsymbol{\beta}}(\boldsymbol{W}_i)$ for $i=1,\ldots,n$. It is also possible to construct a grid specific to each $\boldsymbol{W}_i$ and estimate $\boldsymbol{\beta}(\boldsymbol{W}_i)$ based on that particular grid. In our experience, the resulting estimates based on a single grid (combined with interpolation) and multiple grids (one for each $\boldsymbol{W}_i$) are very similar, given that the single grid is sufficiently dense. For computational efficiency, we suggest using a single grid to estimate $\boldsymbol{\beta}(\boldsymbol{W}_i)$'s.

\subsection{Construction of confidence band for $\boldsymbol{\beta}(\cdot)$}

To construct a confidence interval for $\boldsymbol{\beta}$ at a specific $\boldsymbol{w}$, we can use the asymptotic distribution of $\widehat{\boldsymbol{\beta}}(\boldsymbol{w})$ given in Theorem~1 in Section~3. For the entire function $\boldsymbol{\beta}(\cdot)$, we can construct a simultaneous confidence band using the perturbation method of \cite{tian2005cox}. Suppose that $q=1$, and we are interested in constructing a confidence band for $\boldsymbol{\beta}(w)$ over an interval $[a,b]$. First, we obtain a large-sample approximation to the distribution of
\[
    \mathcal{S}_k=\sup_{w\in[a,b]}\widehat{v}_k(w)\left\vert \widehat{\beta}_k(w)-\beta_{0k}(w)\right\vert\quad \text{for }k=1,\ldots,p,
\]
where $\widehat{v}_k(w)$ is a positive weight function that converges uniformly to a deterministic function.
In particular, we approximate the distribution of the standardized $\mathcal{S}_k$ by considering the stochastic perturbation of the asymptotic linear form of $\widehat{\boldsymbol{\beta}}(w)$ defined by
\begin{align*}
    \widetilde{\boldsymbol{M}}_n(w)=&\frac{1}{n}\sum_{i=1}^n\int_0^\tau\Bigg(K_H(W_i-w)\boldsymbol{X}_i-\sum_{\ell=1}^m K_H(W_i-w_\ell)\overline{\boldsymbol{X}}(t,w)\\
    &-\bigg[\frac{1}{n}\sum_{j=1}^n\int_0^\tau K_H(W_j-w)Y_j(t)\boldsymbol{X}_j\big\{\boldsymbol{Z}_j-\overline{\boldsymbol{Z}}(t)\big\}^\mathrm{T}\,\mathrm{d}t\bigg]\\
    &\times\bigg[\frac{1}{n}\sum_{j=1}^n\int_0^\tau \sum_{\ell=1}^m K_H(W_j-w_\ell)Y_j(t)\big\{\boldsymbol{Z}_j-\overline{\boldsymbol{Z}}(t)\big\}^{\otimes 2}\,\mathrm{d}t\bigg]^{-1}\\
    &\times\sum_{\ell=1}^m K_H(W_i-w_\ell)\big\{\boldsymbol{Z}_i-\overline{\boldsymbol{Z}}(t)\big\}\Bigg)\,\mathrm{d}N_i(t)\psi_i,
\end{align*}
where $\{\psi_i,i=1,\ldots,n\}$ is an iid sample from the standard normal distribution. Note that $\widetilde{\boldsymbol{M}}_n(w)$ is obtained by replacing $M_i(t)$ in a version of $\boldsymbol{M}_n(w)$ (defined in the proof of Theorem~\ref{thm:pointest}) by $N_i(t)\psi_i$ for $i=1,\ldots,n$. We then define
\[
    \widetilde{\mathcal{S}}_k=\sup_{w\in[a,b]}\left\vert \widehat{v}_k(w)\widetilde{M}_{nk}(w)\right\vert,
\]
where $\widetilde{M}_{nk}(w)$ is the $k$th component of the vector $\boldsymbol{D}_n(w)^{-1}\widetilde{\boldsymbol{M}}_n(w)$ and $\boldsymbol{D}_n(w)=n^{-1}\sum_{i=1}^n\int_0^\tau K_H(W_i-w)Y_i(t)\boldsymbol{X}_i^{\otimes 2}$. By repeatedly generating samples of $\{\psi_i,i=1,\ldots,n\}$, we obtain an empirical distribution of $\widetilde{\mathcal{S}}_k$, which can serve as an approximation of the distribution of $\mathcal{S}_k$.

Let $c_{\alpha k}$ be the $(1-\alpha)$th empirical quantile of $\widetilde{\mathcal{S}}_k$, where $0<\alpha<1$. A $(1-\alpha)$ confidence band for $\{\widehat{\beta}_{0k}(w),w\in[a,b]\}$ can be constructed as
\[
    \{\widehat{\beta}_k(w)\pm c_{\alpha k}\widehat{v}_k(w)^{-1},a\leq w\leq b\}.
\]
For $\widehat{v}_k(w)$, we set $\{\widehat{v}_k(w)\}^{-1}$ to be the estimated standard error of $\widehat{\beta}_k(w)$, which is the square root of the $k$th diagonal element of
\begin{align*}
    &\frac{1}{n^2}\boldsymbol{D}_n(w)^{-1}\sum_{i=1}^n\int_0^\tau\Bigg(K_H(W_i-w)\boldsymbol{X}_i-\sum_{\ell=1}^m K_H(W_i-w_\ell)\overline{\boldsymbol{X}}(t,w)\\
    &-\frac{1}{n}\sum_{j=1}^n\int_0^\tau K_H(W_j-w)Y_j(t)\boldsymbol{X}_j\big\{\boldsymbol{Z}_j-\overline{\boldsymbol{Z}}(t)\big\}^\mathrm{T}\,\mathrm{d}t\\
    &\times
    \bigg[\frac{1}{n}\sum_{j=1}^n\int_0^\tau \sum_{\ell=1}^m K_H(W_j-w_\ell)Y_j(t)\big\{\boldsymbol{Z}_j-\overline{\boldsymbol{Z}}(t)\big\}^{\otimes 2}\,\mathrm{d}t\bigg]^{-1}\\
    &\times\sum_{\ell=1}^m K_H(W_i-w_\ell)\big\{\boldsymbol{Z}_i-\overline{\boldsymbol{Z}}(t)\big\}\Bigg)^{\otimes 2}\,\mathrm{d}N_i(t)\boldsymbol{D}_n(w)^{-1}.
\end{align*}

\section{Asymptotic theories}

We establish asymptotic normality of the estimators of the regression parameters and the cumulative baseline hazard function under Conditions~B\ref{cond:1}--B\ref{cond:8} provided in Appendix~\ref{sec:assumptions}. In the theoretical development, we assume that $m\rightarrow\infty$ as $n\rightarrow\infty$, so the grid expands as the sample size increases. The major challenge in the derivations is that $\boldsymbol{b}$ and $\boldsymbol{V}$ increase in dimension as $m\to\infty$, unlike in the local methods where the estimators of $\boldsymbol{\beta}(\boldsymbol{w}_k)$'s can be considered separately. Let $\boldsymbol{\beta}_0\equiv(\boldsymbol{\beta}_0(\boldsymbol{w}_1)^\mathrm{T},\ldots,\boldsymbol{\beta}_0(\boldsymbol{w}_m)^\mathrm{T})^{\mathrm{T}}$, $\boldsymbol{\alpha}_0$, and $\Lambda_0$ denote the true values of $\boldsymbol{\beta}$, $\boldsymbol{\alpha}$, and $\Lambda$, respectively. For simplicity, we assume $\boldsymbol{W}\in[0,1]^q$ with probability one, $\boldsymbol{H}=h\boldsymbol{I}$, and we set $K_h(\boldsymbol{W}-\boldsymbol{w})=\prod_{j=1}^q h^{-1}K(h^{-1}(W_j-w_j))$, where $h$ is a positive tuning parameter.

We first study the asymptotic distribution of $\widehat{\boldsymbol{\beta}}$ at a specific $\boldsymbol{w}_0$ that is in the interior of the support of $\boldsymbol{W}$. Assume that the sets of grid points are nested as $m\rightarrow\infty$ and that each set of grid points includes $\boldsymbol{w}_0$. For $\boldsymbol{w}\notin\{\boldsymbol{w}_1,\ldots,\boldsymbol{w}_m\}$, we define $\widehat{\boldsymbol{\beta}}(\boldsymbol{w})=\widehat{\boldsymbol{\beta}}(\boldsymbol{w}_k)$ for $w_{k,j}\leq w_j<w_{k+1,j}$, where $w_j$ and $w_{k,j}$ represent the $j$th component of $\boldsymbol{w}$ and $\boldsymbol{w}_k$, respectively, for $j=1,\ldots,q$.
\begin{theorem}
\label{thm:pointest}
Under Conditions~B\ref{cond:1}--B\ref{cond:7},
\[
    (nh^q)^{1/2}\big(\widehat{\boldsymbol{\beta}}-\boldsymbol{\beta}_0\big)(\boldsymbol{w}_0)\mathop{\rightarrow}\limits\mathrm{N}(\boldsymbol{0},\boldsymbol{\Sigma}(\boldsymbol{w}_0))
\]
in distribution, where the definition of $\boldsymbol{\Sigma}(\cdot)$ is provided in Appendix~\ref{sec:proof1}.
\end{theorem}

\begin{theorem}
\label{thm:intest}
Under Conditions~B\ref{cond:1}--B\ref{cond:8}, for any bounded function $\boldsymbol{\phi}$,
\[
    n^{1/2}\int_{\mathcal{W}}\boldsymbol{\phi}(\boldsymbol{w})^\mathrm{T}\big\{\widehat{\boldsymbol{\beta}}(\boldsymbol{w})-\boldsymbol{\beta}_0(\boldsymbol{w})\big\}\,\mathrm{d}\boldsymbol{w}\mathop{\rightarrow}\limits\mathrm{N}(\boldsymbol{0},\boldsymbol{\Sigma}_\phi)
\]
in distribution, where the definition of $\boldsymbol{\Sigma}_\phi$ is provided in Appendix~\ref{sec:proof1}.
\end{theorem}

\begin{theorem}
\label{thm:alpha}
Under Conditions~B\ref{cond:1}--B\ref{cond:8},
\[
    n^{1/2}(\widetilde{\boldsymbol{\alpha}}-\boldsymbol{\alpha}_0)\mathop{\rightarrow}\limits\mathrm{N}(\boldsymbol{0},\boldsymbol{\Sigma}_{\widetilde{\alpha}})
\]
in distribution, where the definition of $\boldsymbol{\Sigma}_{\widetilde{\alpha}}$ is provided in Appendix~\ref{sec:proof1}.
\end{theorem}

\begin{theorem}
\label{thm:Lambda}
Under Conditions~B\ref{cond:1}--B\ref{cond:8}, for $t\in[0,\tau]$, $n^{1/2}\{\widehat{\Lambda}(t)-\Lambda_0(t)\}$ is asymptotically normally distributed.
\end{theorem}

The proof of Theorems~\ref{thm:pointest}--\ref{thm:Lambda} are provided in Appendix~\ref{sec:proof1}. Note that $\int_{\mathcal{W}}\boldsymbol{\phi}(\boldsymbol{w})^\mathrm{T}\widehat{\boldsymbol{\beta}}(\boldsymbol{w})\,\mathrm{d}\boldsymbol{w}$ converges at the rate of $n^{1/2}$, whereas $\widehat{\boldsymbol{\beta}}(\boldsymbol{w}_0)$ only converges at the rate of $(nh^q)^{1/2}$; the latter rate is the square root of the size of the local neighborhood at any $w_0$.
As discussed in Section 2.1, a larger grid size $m$ generally improves the approximation of integrals in the estimating equation $\mathcal{V}\beta=\mathcal{C}$.
In the proof of Theorem~\ref{thm:pointest}, we can see that this approximation results in an error term of rate $\max(n^{1/2}h^{-q/2}m^{-1},n^{1/2}h^{q/2+2})$ in the expansion of $\widehat{\boldsymbol{\beta}}(\boldsymbol{w}_0)$, so we require $m\gg(nh^{-q})^{1/2}$, as specified in Condition~B\ref{cond:5}. If $m\gg h^{-3q/2-2}$, then the error term in the expansion of $\widehat{\boldsymbol{\beta}}(\boldsymbol{w}_0)$ is no longer dominated by this approximation error, so further enlarging $m$ does not yield any asymptotic improvement. For the asymptotic normality of $\widehat{\boldsymbol{\beta}}(\boldsymbol{w})$, any positive integer value of $q$ is valid as long as Condition~B\ref{cond:5} is satisfied. By contrast, if the asymptotic normality of $\widetilde{\boldsymbol{\alpha}}$ is required, then Condition~B\ref{cond:8} is needed, which requires $q=1$.

\section{Numerical studies}

\subsection{Simulations}

\begin{figure}[H]
\centering
\includegraphics[scale=0.6]{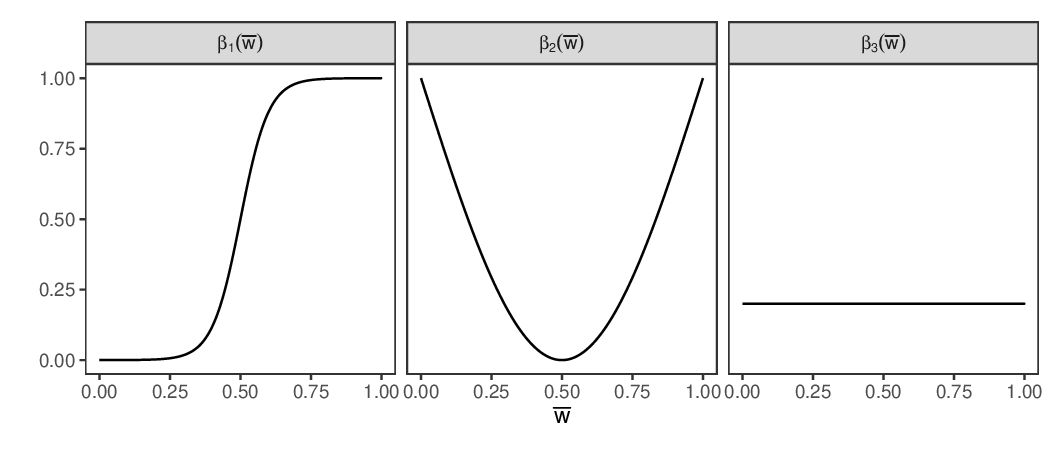}
\caption{True varying coefficient functions in the simulation studies.}
\label{Fig:vc}
\end{figure}

We assess the performance of the proposed methods through simulation studies. We considered a three-dimensional $\boldsymbol{X}$, whose effect on the outcome is modified by a univariate $W$ ($q=1$) or a bivariate $\boldsymbol{W}$ ($q=2$). We considered a bivariate $\boldsymbol{Z}$ that has a constant effect on the hazard rate. The components of $\boldsymbol{W}$, $\boldsymbol{X}$, and $\boldsymbol{Z}$ were generated independently from the Uniform$(0,1)$ distribution. We generated the outcome variable from the following hazard function:
\begin{align*}
    \lambda(t\mid\boldsymbol{W}=\boldsymbol{w},\boldsymbol{X},\boldsymbol{Z})&=\lambda(t)+\beta_1(\boldsymbol{w})X_1+\beta_2(\boldsymbol{w})X_2+\beta_3(\boldsymbol{w})X_3+\alpha_1 Z_1+\alpha_2 Z_2\\
    &=t+\frac{1}{1+e^{-20(\overline{w}-0.5)}}X_1+\{1-\sin(\pi\overline{w})\}X_2+0.2X_3+0.2Z_1+0.2Z_2,
\end{align*}
where $\overline{w}$ is the sample mean of the components of $\boldsymbol{w}$. The functions $\beta_1(\boldsymbol{w})$, $\beta_2(\boldsymbol{w})$, and $\beta_3(\boldsymbol{w})$ are plotted in Fig.~\ref{Fig:vc}. We set the censoring time to follow an exponential distribution with the mean chosen to yield a censoring rate of about 30\%. We considered sample sizes of $n=200$, 500, and 1000.

In addition to the proposed method, we considered two alternative methods: the additive hazards model on linear predictors $\boldsymbol{X}$ and $\boldsymbol{Z}$, and the varying coefficient additive hazards model, estimated using the local kernel method. For the local method, we first consider $\boldsymbol{\alpha}$ as varying and estimate $(\boldsymbol{\alpha},\boldsymbol{\beta}(\boldsymbol{W}_i))$ for each $\boldsymbol{W}_i$ ($i=1,\ldots,n$). Following \cite{yin2008partially}, we set the final estimator of $\boldsymbol{\alpha}$ as a weighted average across the estimators $\widehat{\boldsymbol{\alpha}}(\boldsymbol{W}_i)$ at each $\boldsymbol{W}_i$. For the proposed method, under a univariate $W$, we obtained $\widehat{\boldsymbol{\beta}}$ at an evenly spaced grid over $[0,1]$ with size 5, 9, or 13. Under a bivariate $\boldsymbol{W}$, we considered an evenly spaced grid over $[0,1]$ with size 5 in each dimension.  We used linear interpolation to estimate $\boldsymbol{\beta}$ at points off the grid.

\begin{figure}
\centering
\includegraphics[scale=0.6]{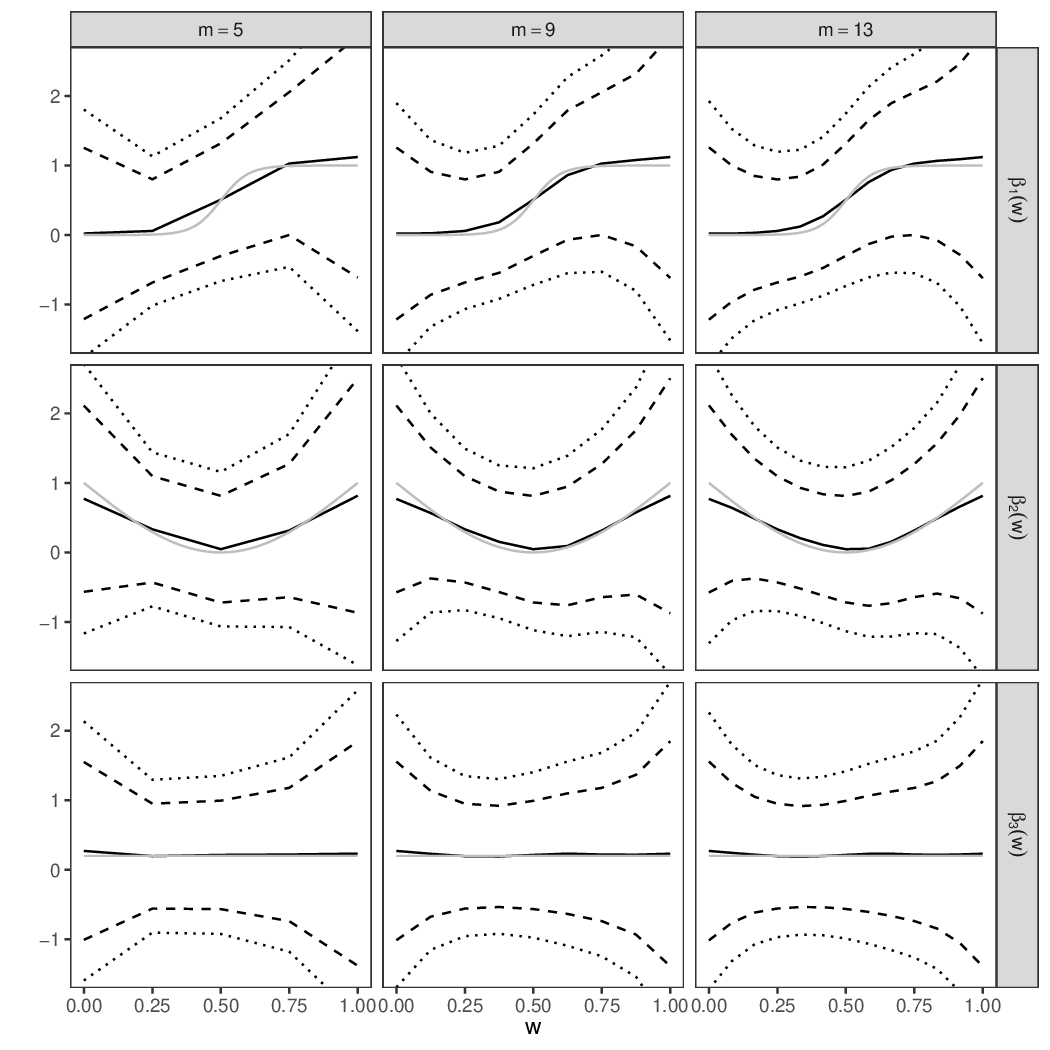}
\caption{The estimated coefficient functions (solid in black), true coefficient functions (solid in gray), 95\% simultaneous confidenece interval (dotted), and 95\% pointwise confidence intervals (dashed) for $q=1$ under $n=200$.}
\label{Fig:beta est n200}
\end{figure}

\begin{figure}
\centering
\includegraphics[scale=0.6]{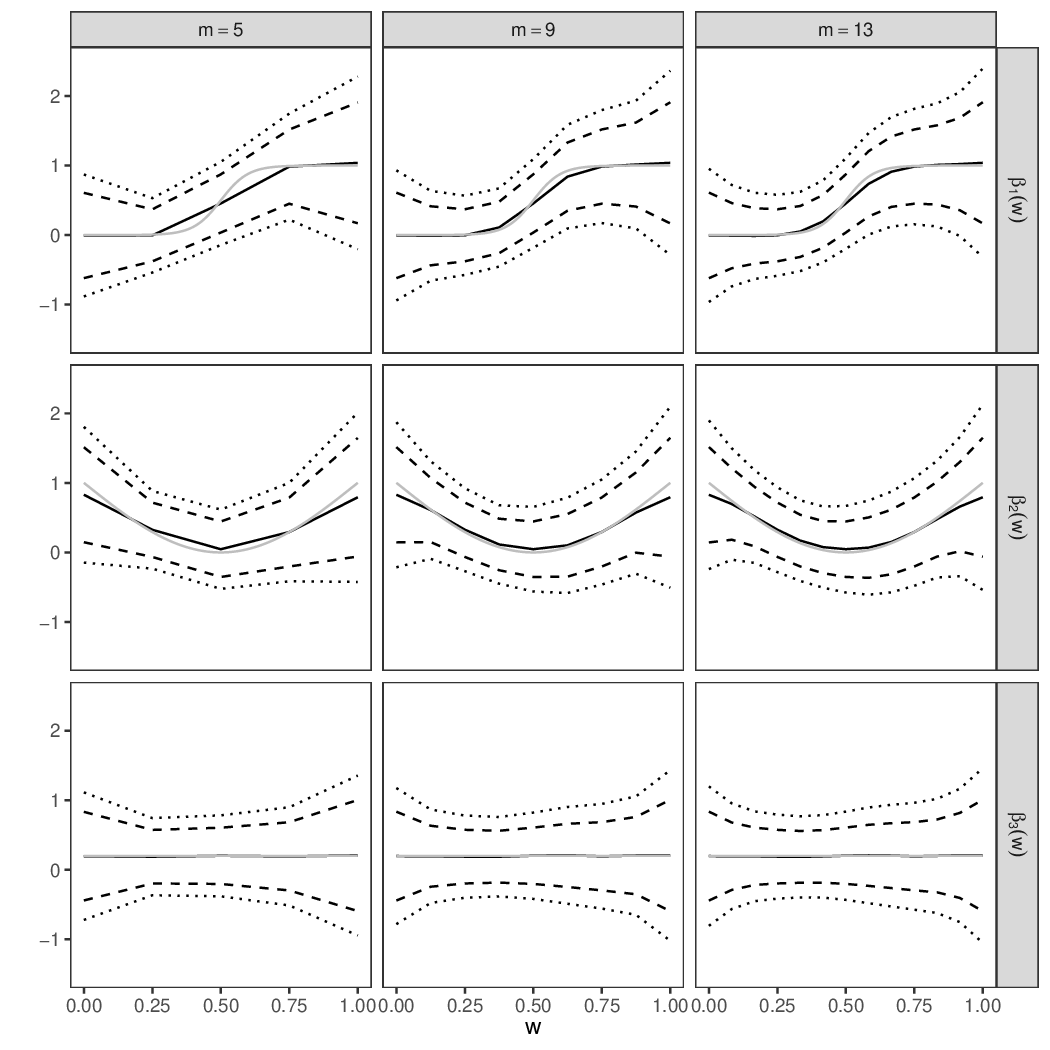}
\caption{The estimated coefficient functions (solid in black), true coefficient functions (solid in gray), 95\% simultaneous confidenece interval (dotted), and 95\% pointwise confidence intervals (dashed) for $q=1$ under $n=1000$.}
\label{Fig:beta est n1000}
\end{figure}

\begin{figure}
\centering
\includegraphics[scale=0.6]{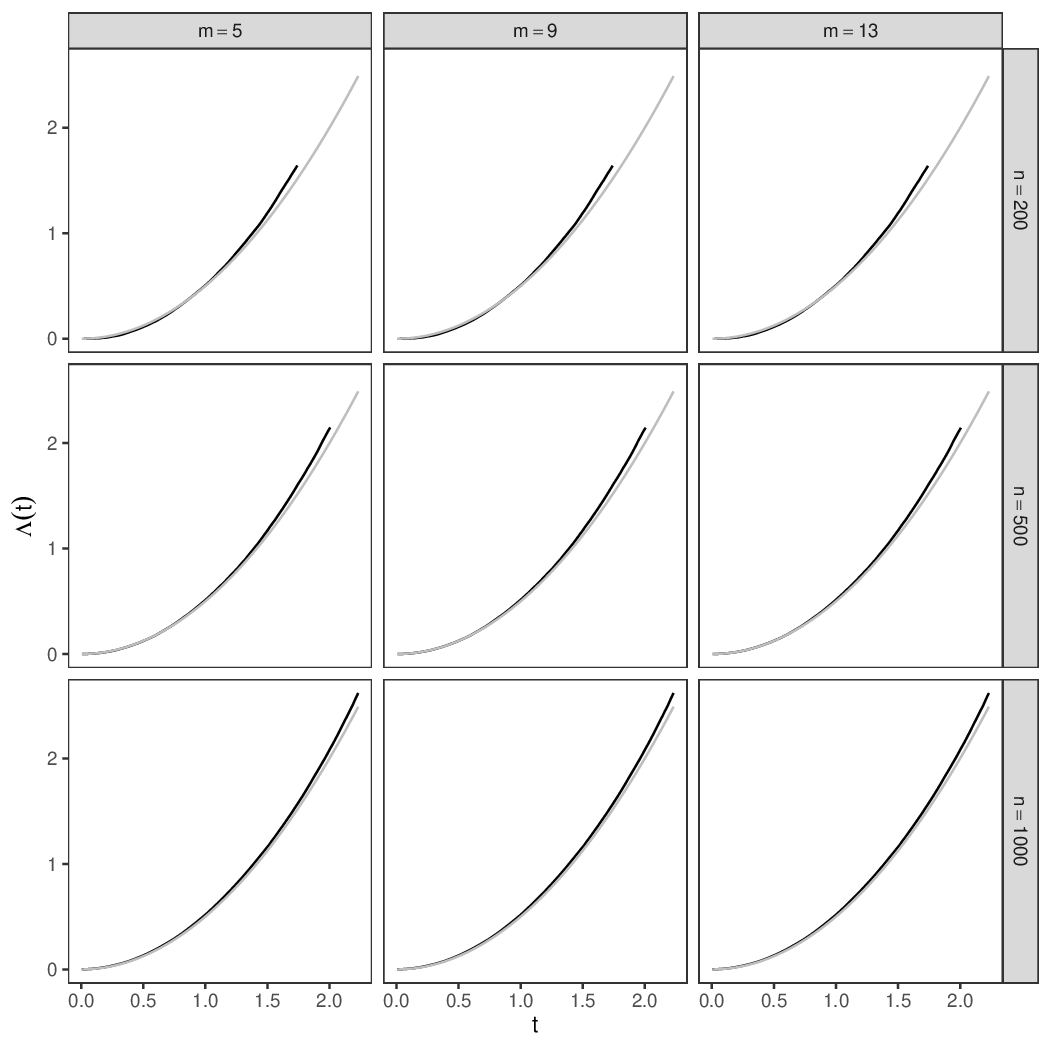}
\caption{The estimated cumulative baseline hazard function (black) and the true cumulative baseline hazard function (gray) for $q=1$.}
\label{Fig:Lambda0 est}
\end{figure}

\begin{figure}
\centering
\includegraphics[scale=0.6]{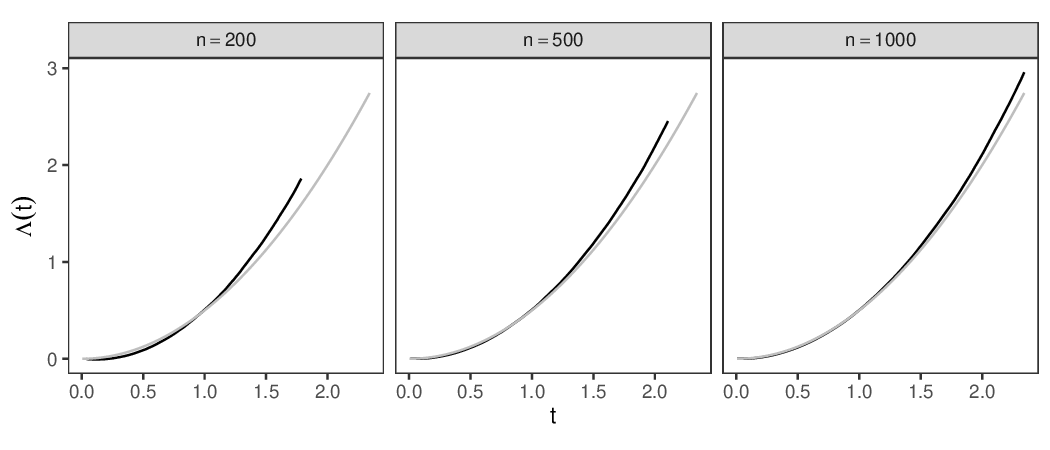}
\caption{The estimated cumulative baseline hazard function (black) and the true cumulative baseline hazard function (gray) for $q=2$.}
\label{Fig:Lambda0 est q2}
\end{figure}

The performance of each method is assessed using the following measures. We report the mean-squared error, defined as $\mathrm{E}[\{\widehat{\boldsymbol{\beta}}(\boldsymbol{W})-\boldsymbol{\beta}_{0}(\boldsymbol{W})\}^{\mathrm{T}}\boldsymbol{X}+(\widehat{\boldsymbol{\alpha}}-\boldsymbol{\alpha}_{0})^{\mathrm{T}}\boldsymbol{Z}]^{2}$, where $\widehat{\boldsymbol{\beta}}(\boldsymbol{W})$ and $\boldsymbol{\beta}_{0}(\boldsymbol{W})$ denote the estimated and true values of $\boldsymbol{\beta}$ evaluated at $\boldsymbol{W}$, respectively, and $\widehat{\boldsymbol{\alpha}}$ and $\boldsymbol{\alpha}_{0}$ denote the estimated and true values of $\boldsymbol{\alpha}$, respectively. Also, we compute the concordance index between  $\widehat{\boldsymbol{\beta}}(\boldsymbol{W})^{\mathrm{T}}\boldsymbol{X}+\widehat{\boldsymbol{\alpha}}^{\mathrm{T}}\boldsymbol{Z}$ and the event time on an independent sample. We present the sample standard deviations of the estimates and the average estimated standard errors, computed based on asymptotic approximation. We provide the coverage rates of the 95\% confidence intervals. We considered 500 simulation replicates. The results for $q=1$ are summarized in Tables~\ref{Tab:simulation results(q1)}--\ref{Tab:CR q1}. Figures~\ref{Fig:beta est n200}--\ref{Fig:beta est n1000} show the average estimated values of $\beta_1$, $\beta_2$, and $\beta_3$. The estimated cumulative baseline hazard functions under $q=1$ are plotted in Fig.~\ref{Fig:Lambda0 est}. The simulation results for $q=2$ are presented in Tables~\ref{Tab:simulation results(q2)}--\ref{Tab:CR q2} and Fig.~\ref{Fig:Lambda0 est q2}.

\begin{table}
\centering
\caption{MSE and C-index for $q=1$}
\begin{threeparttable}
\begin{tabular}{rrrrr}
 \\
& Method & $n=200$ & $n=500$ & $n=1000$\\
MSE & Constant & 0.317 & 0.192 & 0.157\\
& Local & 0.579 & 0.228 & 0.133\\
& Proposed ($m=5$) & 0.303 & 0.121 & 0.066\\
& Proposed ($m=9$) & 0.309 & 0.123 & 0.064\\
& Proposed ($m=13$) & 0.312 & 0.125 & 0.066\\
C-index & Constant & 0.532 & 0.541 & 0.545\\
& Local & 0.548 & 0.565 & 0.575\\
& Proposed ($m=5$) & 0.568 & 0.582 & 0.590\\
& Proposed ($m=9$) & 0.568 & 0.582 & 0.590\\
& Proposed ($m=13$) & 0.568 & 0.582 & 0.591
\end{tabular}
\label{Tab:simulation results(q1)}
\begin{tablenotes}
MSE, mean-squared error; C-index, concordance index; Constant, regression model on linear predictors $\boldsymbol{X}$ and $\boldsymbol{Z}$; Local, the varying coefficient model where the local kernel estimation method is adopted; Proposed, the varying coefficient model with the proposed estimation method.
\end{tablenotes}
\end{threeparttable}
\end{table}

\begin{table}\centering
\caption{Estimation of regression coefficients for $q=1$}
\begin{threeparttable}
\begin{tabular}{ccrrrrrrrr}
 \\
& & \multicolumn{4}{c}{$\alpha_1$} & \multicolumn{4}{c}{$\alpha_2$}\\
n & $m$ & Bias & SD & SE & CR & Bias & SD & SE & CR\\
200 & 5 & 0.016 & 0.380 & 0.350 & 0.962 & 0.028 & 0.360 & 0.349 & 0.970\\
& 9 & 0.017 & 0.381 & 0.350 & 0.962 & 0.030 & 0.362 & 0.349 & 0.972\\
& 13 & 0.017 & 0.382 & 0.350 & 0.962 & 0.030 & 0.363 & 0.349 & 0.970\\
500 & 5 & 0.015 & 0.215 & 0.219 & 0.978 & $-$0.012 & 0.216 & 0.218 & 0.976\\
& 9 & 0.016 & 0.216 & 0.219 & 0.976 & $-$0.012 & 0.216 & 0.218 & 0.976\\
& 13 & 0.016 & 0.216 & 0.219 & 0.976 & $-$0.011 & 0.216 & 0.218 & 0.976\\
1000 & 5 & $-$0.011 & 0.154 & 0.154 & 0.976 & 0.000 & 0.157 & 0.154 & 0.970\\
& 9 & $-$0.011 & 0.155 & 0.154 & 0.976 & 0.001 & 0.158 & 0.154 & 0.974\\
& 13 & $-$0.011 & 0.155 & 0.154 & 0.978 & 0.001 & 0.159 & 0.154 & 0.974
\end{tabular}
\label{Tab:alpha}
\begin{tablenotes}
SD, sample standard deviations of the estimates; SE, estimated standard errors based on asymptotic approximation; CR, coverage rates of the 95\% confidence intervals.
\end{tablenotes}
\end{threeparttable}
\end{table}

\begin{table}
\centering
\caption{Bias and standard deviation (in parentheses) of the estimated covariate effects for $q=1$}
\begin{threeparttable}
\begin{tabular}{cccrrrr}
\\
$n$ & & $w$ & Local & \multicolumn{3}{c}{Proposed}\\
& & & & $m=5$ & $m=9$ & $m=13$\\
200 & $\alpha_1$ & & 0.010 (0.367) & 0.017 (0.380) & 0.016 (0.377) & 0.017 (0.377)\\
& $\alpha_2$ & & 0.014 (0.354) & 0.029 (0.359) & 0.029 (0.360) & 0.029 (0.361)\\
& $\beta_1(w)$ & 0.2 & 0.028 (0.542) & 0.050 (0.462) & 0.043 (0.474) & 0.041 (0.484)\\
& & 0.4 & 0.113 (0.510) & 0.210 (0.418) & 0.130 (0.461) & 0.122 (0.470)\\
& & 0.6 & $-$0.096 (0.617) & $-$0.164 (0.468) & $-$0.089 (0.553) & $-$0.084 (0.562)\\
& & 0.8 & 0.046 (0.735) & 0.050 (0.600) & 0.051 (0.624) & 0.054 (0.639)\\
& $\beta_2(w)$ & 0.2 & $-$0.007 (0.590) & 0.008 (0.488) & 0.014 (0.509) & 0.012 (0.524)\\
& & 0.4 & 0.080 (0.553) & 0.112 (0.431) & 0.082 (0.474) & 0.078 (0.482)\\
& & 0.6 & 0.005 (0.586) & 0.105 (0.459) & 0.035 (0.522) & 0.025 (0.531)\\
& & 0.8 & $-$0.003 (0.741) & 0.003 (0.587) & 0.007 (0.605) & 0.007 (0.619)\\
& $\beta_3(w)$ & 0.2 & $-$0.005 (0.569) & 0.011 (0.492) & 0.010 (0.506) & 0.009 (0.515)\\
& & 0.4 & $-$0.004 (0.538) & 0.007 (0.444) & $-$0.003 (0.478) & $-$0.004 (0.488)\\
& & 0.6 & 0.025 (0.604) & 0.017 (0.458) & 0.029 (0.524) & 0.030 (0.533)\\
& & 0.8 & 0.007 (0.717) & 0.023 (0.588) & 0.019 (0.609) & 0.017 (0.625)\\
500 & $\alpha_1$ & & 0.012 (0.217) & 0.014 (0.217) & 0.015 (0.216) & 0.015 (0.216)\\
& $\alpha_2$ & & $-$0.017 (0.212) & $-$0.012 (0.219) & $-$0.013 (0.215) & $-$0.014 (0.215)\\
& $\beta_1(w)$ & 0.2 & 0.015 (0.356) & 0.024 (0.294) & 0.022 (0.298) & 0.020 (0.307)\\
& & 0.4 & 0.058 (0.360) & 0.179 (0.270) & 0.092 (0.308) & 0.083 (0.318)\\
& & 0.6 & $-$0.091 (0.408) & $-$0.182 (0.289) & $-$0.104 (0.345) & $-$0.100 (0.352)\\
& & 0.8 & 0.064 (0.487) & 0.046 (0.397) & 0.046 (0.418) & 0.051 (0.435)\\
& $\beta_2(w)$ & 0.2 & $-$0.006 (0.374) & $-$0.008 (0.304) & 0.001 (0.316) & 0.000 (0.328)\\
& & 0.4 & 0.006 (0.342) & 0.095 (0.265) & 0.043 (0.290) & 0.038 (0.299)\\
& & 0.6 & 0.030 (0.398) & 0.092 (0.263) & 0.035 (0.326) & 0.030 (0.337)\\
& & 0.8 & 0.009 (0.448) & $-$0.016 (0.359) & 0.002 (0.372) & 0.002 (0.387)\\
& $\beta_3(w)$ & 0.2 & 0.012 (0.359) & 0.021 (0.296) & 0.016 (0.301) & 0.015 (0.310)\\
& & 0.4 & 0.001 (0.334) & 0.015 (0.261) & 0.019 (0.291) & 0.019 (0.298)\\
& & 0.6 & 0.012 (0.423) & 0.002 (0.285) & 0.005 (0.344) & 0.006 (0.351)\\
& & 0.8 & $-$0.007 (0.492) & $-$0.014 (0.373) & $-$0.021 (0.392) & $-$0.021 (0.410)\\
1000 & $\alpha_1$ & & $-$0.015 (0.154) & $-$0.008 (0.156) & $-$0.011 (0.155) & $-$0.012 (0.155)\\
& $\alpha_2$ & & $-$0.003 (0.154) & $-$0.001 (0.161) & 0.001 (0.157) & 0.001 (0.157)\\
& $\beta_1(w)$ & 0.2 & $-$0.011 (0.268) & $-$0.007 (0.198) & $-$0.009 (0.206) & $-$0.011 (0.217)\\
& & 0.4 & 0.036 (0.257) & 0.152 (0.204) & 0.059 (0.226) & 0.048 (0.233)\\
& & 0.6 & $-$0.106 (0.322) & $-$0.214 (0.215) & $-$0.117 (0.265) & $-$0.109 (0.273)\\
& & 0.8 & 0.007 (0.353) & $-$0.002 (0.266) & $-$0.001 (0.272) & 0.002 (0.290)\\
& $\beta_2(w)$ & 0.2 & 0.029 (0.273) & 0.014 (0.217) & 0.027 (0.221) & 0.027 (0.232)\\
& & 0.4 & 0.036 (0.259) & 0.109 (0.190) & 0.052 (0.215) & 0.046 (0.222)\\
& & 0.6 & 0.025 (0.297) & 0.096 (0.202) & 0.042 (0.245) & 0.036 (0.250)\\
& & 0.8 & $-$0.007 (0.352) & $-$0.019 (0.269) & $-$0.007 (0.273) & $-$0.008 (0.288)\\
& $\beta_3(w)$ & 0.2 & $-$0.003 (0.277) & $-$0.010 (0.213) & $-$0.009 (0.220) & $-$0.008 (0.233)\\
& & 0.4 & $-$0.012 (0.264) & $-$0.004 (0.195) & $-$0.009 (0.227) & $-$0.009 (0.235)\\
& & 0.6 & 0.003 (0.303) & $-$0.002 (0.214) & 0.006 (0.247) & 0.007 (0.253)\\
& & 0.8 & $-$0.002 (0.377) & $-$0.004 (0.282) & $-$0.002 (0.290) & $-$0.002 (0.307)
\end{tabular}
\label{Tab:bias and sd(q1)}
\begin{tablenotes}
Local, the varying coefficient model where the local kernel estimation method is adopted; Proposed, the varying coefficient model with the proposed estimation method.
\end{tablenotes}
\end{threeparttable}
\end{table}

\begin{table}
\centering
\caption{Coverage rates of the varying coefficients for $q=1$}
\begin{threeparttable}
\begin{tabular}{ccrrr}
 \\
$n$ & $m$ & $\beta_1(w)$ & $\beta_2(w)$ & $\beta_3(w)$\\
200 & 5 & 0.982 & 0.972 & 0.982\\
& 9 & 0.974 & 0.964 & 0.974\\
& 13 & 0.974 & 0.964 & 0.970\\
500 & 5 & 0.988 & 0.978 & 0.978\\
& 9 & 0.978 & 0.980 & 0.972\\
& 13 & 0.980 & 0.978 & 0.978\\
1000 & 5 & 0.984 & 0.974 & 0.972\\
& 9 & 0.980 & 0.978 & 0.982\\
& 13 & 0.982 & 0.978 & 0.980
\end{tabular}
\label{Tab:CR q1}
\end{threeparttable}
\end{table}

\begin{table}
\centering
\caption{MSE and C-index for $q=2$}
\begin{threeparttable}
\begin{tabular}{rrrrr}
 \\
& Method & $n=200$ & $n=500$ & $n=1000$\\
MSE & Constant & 0.285 & 0.154 & 0.114\\
& Local & 1.521 & 0.626 & 0.341\\
& Proposed & 0.514 & 0.183 & 0.094\\
C-index & Constant & 0.528 & 0.537 & 0.541\\
& Local & 0.524 & 0.534 & 0.542\\
& Proposed & 0.545 & 0.559 & 0.566
\end{tabular}
\label{Tab:simulation results(q2)}
\begin{tablenotes}
MSE, mean-squared error; C-index, concordance index; Constant, regression model on linear predictors $\boldsymbol{X}$ and $\boldsymbol{Z}$; Local, the varying coefficient model where the local kernel estimation method is adopted; Proposed, the varying coefficient model with the proposed estimation method.
\end{tablenotes}
\end{threeparttable}
\end{table}

\begin{figure}
\centering
\includegraphics[scale=0.6]{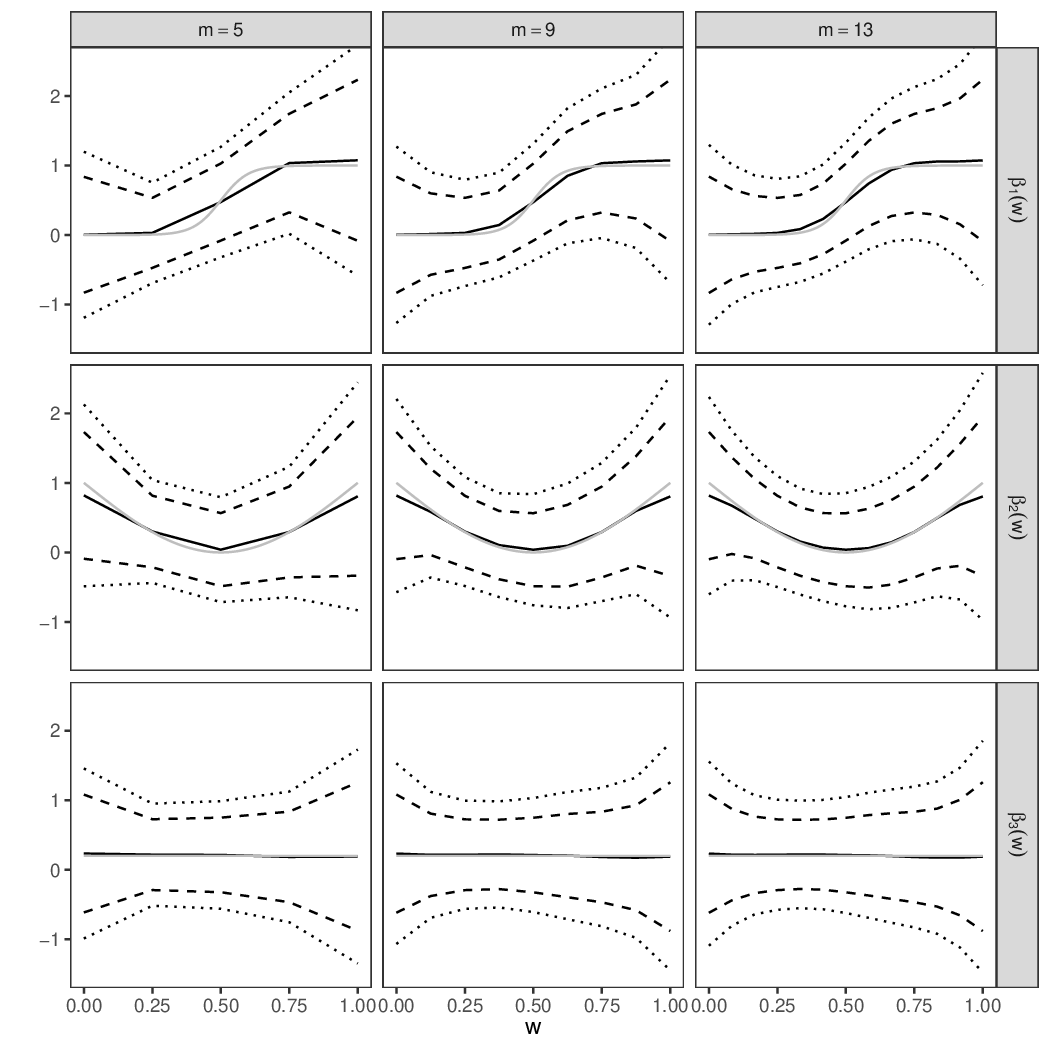}
\caption{The estimated coefficient functions (solid in black), true coefficient functions (solid in gray), 95\% simultaneous confidenece interval (dotted), and 95\% pointwise confidence intervals (dashed) for $q=1$ under $n=500$.}
\end{figure}

\begin{table}
\centering
\caption{Estimation of regression coefficients for $q=2$}
\begin{threeparttable}
\begin{tabular}{ccrrrrrrr}
 \\
& \multicolumn{4}{c}{$\alpha_1=0.2$} & \multicolumn{4}{c}{$\alpha_2=0.2$}\\
n & Bias & SD & SE & CR & Bias & SD & SE & CR\\
200 & 0.028 & 0.385 & 0.334 & 0.956 & 0.031 & 0.374 & 0.335 & 0.960\\
500 & 0.013 & 0.217 & 0.208 & 0.964 & $-$0.008 & 0.227 & 0.209 & 0.966\\
1000 & 0.013 & 0.154 & 0.147 & 0.968 & 0.002 & 0.150 & 0.147 & 0.968
\end{tabular}
\begin{tablenotes}
SD, sample standard deviations of the estimates; SE, estimated standard errors based on asymptotic approximation; CR, coverage rates of the 95\% confidence intervals.
\end{tablenotes}
\end{threeparttable}
\end{table}

\begin{table}
\centering
\caption{Bias and standard deviation (in parentheses) of the estimated covariate effects for $q=2$}
\begin{threeparttable}
\begin{tabular}{cccrr}
\\
$n$ & & $\boldsymbol{w}$ & Local & Proposed\\
200 & $\alpha_1$ & & $-$0.001 (0.356) & 0.026 (0.381)\\
& $\alpha_2$ & & 0.004 (0.358) & 0.029 (0.372)\\
& $\beta_1(\boldsymbol{w})$ & (0.25,0.25) & 0.052 (0.800) & 0.066 (0.685)\\
& & (0.25,0.75) & 0.048 (0.851) & 0.053 (0.801)\\
& & (0.75,0.25) & 0.011 (0.909) & 0.055 (0.743)\\
& & (0.75,0.75) & 0.098 (1.104) & 0.094 (0.920)\\
& $\beta_2(\boldsymbol{w})$ & (0.25,0.25) & 0.013 (0.726) & 0.043 (0.664)\\
& & (0.25,0.75) & 0.053 (0.841) & 0.000 (0.727)\\
& & (0.75,0.25) & $-$0.029 (0.852) & 0.067 (0.720)\\
& & (0.75,0.75) & 0.133 (1.157) & 0.131 (0.979)\\
& $\beta_3(\boldsymbol{w})$ & (0.25,0.25) & $-$0.015 (0.851) & 0.011 (0.713)\\
& & (0.25,0.75) & 0.002 (0.835) & 0.101 (0.770)\\
& & (0.75,0.25) & 0.079 (0.875) & 0.010 (0.741)\\
& & (0.75,0.75) & $-$0.017 (1.074) & $-$0.001 (0.855)\\
500 & $\alpha_1$ & & $-$0.005 (0.214) &  0.013 (0.218)\\
& $\alpha_2$ & & $-$0.017 (0.221) & $-$0.007 (0.228)\\
& $\beta_1(\boldsymbol{w})$ & (0.25,0.25) & $-$0.032 (0.528) & $-$0.013 (0.446)\\
& & (0.25,0.75) & 0.018 (0.559) & 0.005 (0.529)\\
& & (0.75,0.25) & $-$0.037 (0.605) & 0.018 (0.494)\\
& & (0.75,0.75) & 0.060 (0.752) & 0.047 (0.647)\\
& $\beta_2(\boldsymbol{w})$ & (0.25,0.25) & $-$0.029 (0.563) & $-$0.011 (0.491)\\
& & (0.25,0.75) & 0.029 (0.593) & 0.012 (0.515)\\
& & (0.75,0.25) & $-$0.032 (0.604) & 0.025 (0.493)\\
& & (0.75,0.75) & 0.017 (0.695) & 0.024 (0.595)\\
& $\beta_3(\boldsymbol{w})$ & (0.25,0.25) & 0.039 (0.563) & 0.064 (0.505)\\
& & (0.25,0.75) & 0.021 (0.579) & 0.066 (0.546)\\
& & (0.75,0.25) & 0.023 (0.628) & 0.018 (0.510)\\
& & (0.75,0.75) & 0.066 (0.708) & 0.060 (0.599)\\
1000 & $\alpha_1$ & & 0.006 (0.154) & 0.014 (0.158)\\
& $\alpha_2$ & & $-$0.007 (0.148) & 0.003 (0.152)\\
& $\beta_1(\boldsymbol{w})$ & (0.25,0.25) & 0.034 (0.395) & 0.022 (0.336)\\
& & (0.25,0.75) & 0.001 (0.437) & $-$0.024 (0.388)\\
& & (0.75,0.25) & $-$0.024 (0.466) & 0.006 (0.370)\\
& & (0.75,0.75) & 0.043 (0.578) & 0.048 (0.510)\\
& $\beta_2(\boldsymbol{w})$ & (0.25,0.25) & 0.023 (0.423) & 0.013 (0.376)\\
& & (0.25,0.75) & $-$0.001 (0.455) & 0.024 (0.372)\\
& & (0.75,0.25) & 0.023 (0.432) & 0.002 (0.398)\\
& & (0.75,0.75) & 0.023 (0.592) & 0.023 (0.480)\\
& $\beta_3(\boldsymbol{w})$ & (0.25,0.25) & 0.023 (0.431) & 0.014 (0.368)\\
& & (0.25,0.75) & 0.012 (0.446) & 0.043 (0.391)\\
& & (0.75,0.25) & 0.042 (0.448) & 0.012 (0.382)\\
& & (0.75,0.75) & $-$0.010 (0.547) & $-$0.007 (0.462)
\end{tabular}
\label{Tab:bias and se(q2)}
\begin{tablenotes}
Local, the varying coefficient model where the local kernel estimation method is adopted; Proposed, the varying coefficient model with the proposed estimation method.
\end{tablenotes}
\end{threeparttable}
\end{table}

\begin{table}
\centering
\caption{Coverage rates of the varying coefficients for $q=2$}
\begin{threeparttable}
\begin{tabular}{crrr}
 \\
$n$ & $\beta_1(\boldsymbol{w})$ & $\beta_2(\boldsymbol{w})$ & $\beta_3(\boldsymbol{w})$\\
200 & 0.990 & 0.992 & 0.994\\
500 & 0.992 & 0.998 & 0.994\\
1000 & 0.992 & 0.990 & 0.996
\end{tabular}
\label{Tab:CR q2}
\end{threeparttable}
\end{table}

Overall, the proposed method yields higher prediction accuracy than the alternatives. In most simulation settings, the proposed method yields superior prediction performance, with the lowest mean-squared error and highest concordance index values. The proposed method outperforms the local method, because the proposed method accounts for the fact that all subjects share the same baseline hazard function. Also, it can be seen that the proposed method yields much higher predictive power than the local method when the sample size is small. This demonstrates the importance of utilizing the information of the baseline hazard function from all subjects especially when the sample size is limited. The biases under the two kernel-based estimation methods are similar, but the standard errors of the estimates under the proposed method are smaller than those under the local method. 
Also, the biases for the grid points near the boundaries are relatively large due to limited samples. In the cases with small samples, this may lead to larger mean-squared error. The biases reduce when the sample size increases.

For $m=5$, 9, and 13, the mean-squared error and concordance index values are almost the same. As shown in Fig.~\ref{Fig:beta est n1000}, when the number of grid points increases, the biases reduce as more interpolation grids can generally capture the curvilinear structure of the true function. Nevertheless, the improvement in the estimation accuracy is not substantial for $\beta_1$, which is a monotone function. If the true function is steadily increasing or decreasing, then a small number of grid points may be sufficient to capture the variation in the varying coefficient, and the prediction performance is not sensitive to the number of grids. For $\beta_2$, there is a considerable improvement in bias when we increase the number of grids from 5 to 13. If the true function has (multiple) turns, then more grid points are required to capture the variation of the varying coefficient over different values of $w$.

\subsection{Real data analysis}

We demonstrate the application of the proposed method using a dataset of kidney renal clear cell carcinoma (KIRC) patients from The Cancer Genome Atlas (TCGA, \citeyear{tcga2013comprehensive}), available at \texttt{https://gdac.broadinstitute.org}. In particular, we investigated the effects of protein expressions, gene expressions, and clinical variables on time to death since initial diagnosis, allowing for interactions between protein and gene expressions. We fit the proposed model for TSC-mTOR pathway that have known role in cancer progression \citep{akbani2014pan}. We set $\boldsymbol{X}$ to be the expressions of 4E-BP1\_pS65, 4E-BP1\_pT37\_T46, and 4E-BP1\_pT70 and set $W$ to be the expression of EIF4EBP1. This formulation allows the effects of protein expressions to be modified by gene expressions. The set of linear predictors $\boldsymbol{Z}$ consists of age and gender (0 for female and 1 male), which allows linear effects of clinical variables on overall survival. After removing subjects with missing data, the sample size is 475. The median time to censoring or death is 3.22 years, and the censoring rate is 65.26\%. We set an evenly-spaced grid points with size 50.

\begin{figure}[H]
\centering
\includegraphics[scale=0.6]{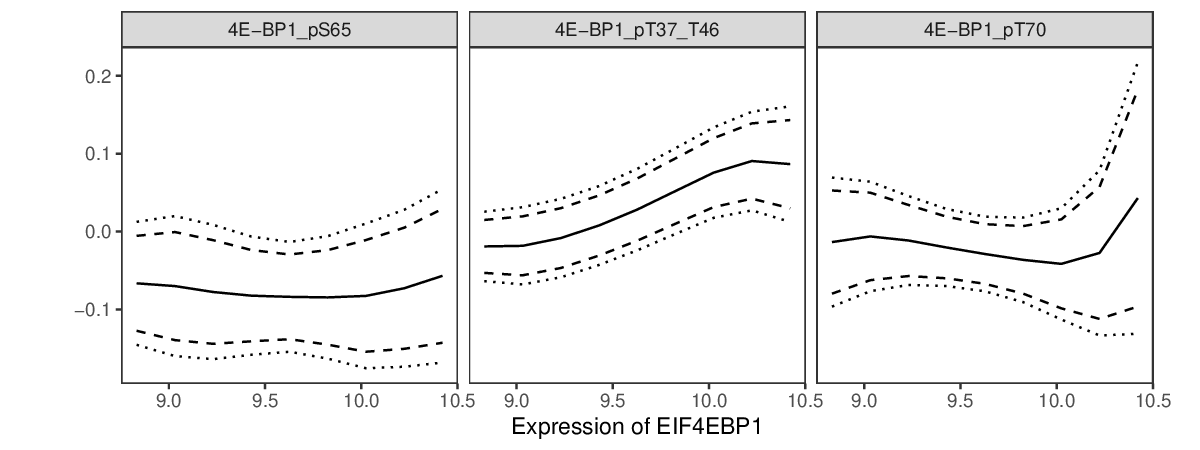}
\caption{The estimated coefficient functions (solid), 95\% simultaneous confidenece interval (dotted), and 95\% pointwise confidence intervals (dashed) for the KIRC analysis.}
\label{Fig:KIRC}
\end{figure}

For comparison, we considered two alternative methods: the additive hazards model on linear predictors $\boldsymbol{X}$ and $\boldsymbol{Z}$, and the varying coefficient additive hazards model, on which the local kernel estimation method is adopted. The estimated constant coefficients under different methods are shown in Table~\ref{Tab:KIRC} and the estimated varying coefficients are shown in Fig.~\ref{Fig:KIRC}. The effects of 4E-BG1\_pS65 and 4E-BP1\_pT70 tend to remain constant. The effect of 4E-BP1\_pT37\_T46 appears to increase as the expression of EIF4EBP1 increases, which suggests a potential varying covariate effect.

\begin{table}[H]
\centering
\caption{Estimated constant covariate effect and standard deviation (in parentheses) for the KIRC analysis}
\begin{threeparttable}
\begin{tabular}{lrrr}
 \\
Variable & Constant & Local & Proposed\\
4E-BP1\_pS65 & $-$0.079 (0.022) & - & -\\
4E-BP1\_pT37\_T46 & 0.045 (0.015) & - & -\\
4E-BP1\_pT70 & $-$0.003 (0.020) & - & -\\
Age & 0.003 (0.001) & 0.000 (0.000) & 0.003 (0.001)\\
Gender & 0.003 (0.016) & $-$0.015 (0.004) & 0.004 (0.016)
\end{tabular}
\label{Tab:KIRC}
\begin{tablenotes}
Constant, regression model on linear predictors $\boldsymbol{X}$ and $\boldsymbol{Z}$; Local, the varying coefficient model where the local kernel estimation method is adopted; Proposed, the varying coefficient model with the proposed estimation method.
\end{tablenotes}
\end{threeparttable}
\end{table}

\section{Discussion}

In this paper, we study kernel-based estimation methods for the varying coefficient additive hazards models. Many existing kernel-based methods for varying coefficient survival models are direct extensions of methods for corresponding models without varying coefficients that incorporate a kernel weight to each subject in existing estimators. Such local methods suffer  efficiency loss, because they do not account for the fact that there are some nuisance parameters (in the cases of the proportional hazards and additive hazards models, the baseline hazard function) shared among all subjects. The proposed global kernel estimation method, by contrast, uses all subjects to profile out the nuisance parameters and thus has higher efficiency.

In the literature, estimation methods for varying coefficient models are mainly based on the kernel smoothing or spline approximation. Although the two approaches may yield similar empirical results for univariate varying coefficient functions, the kernel approach is more flexible under a multivariate $\boldsymbol{W}$. In particular, for kernel methods, we simply need to define the distance between two values of $\boldsymbol{W}$, which is not necessarily the Euclidean distance. This distance can then be used to determine the subject weights in estimation, and $\boldsymbol{\beta}(\boldsymbol{W})$ at two $\boldsymbol{W}$ values close to each other (according to the distance) would be encouraged to be similar. By contrast, it would be more challenging to define (multivariate) spline functions that encourage $\boldsymbol{\beta}(\boldsymbol{w}_1)$ and $\boldsymbol{\beta}(\boldsymbol{w}_2)$ to be similar for $\boldsymbol{w}_1$ and $\boldsymbol{w}_2$ close to each other under some non-Euclidean distance. 

The proposed estimator is motivated by first considering a standard estimator of the proposed method under a discrete $\boldsymbol{W}$ and then ``kernelizing'' it, that is, suitably incorporating kernel weights to the estimator. This general approach can be applied to other varying coefficient models that contain parameters not varying with $\boldsymbol{W}$, such as the (varying coefficient) Cox proportional hazards model or transformation model \citep{cheng1995analysis}, to yield global estimators. In fact, such an approach for the varying coefficient Cox model would yield an estimator that is asymptotically equivalent to that of \cite{chen2012efficient}, although \cite{chen2012efficient} proposed an alternative computation method that iteratively updates (the counterparts of) $\boldsymbol{\beta}(\boldsymbol{W}_i)$ and $\boldsymbol{\alpha}$ for $i=1,\ldots,n$.

The current work can be extended in several directions. First, the proposed method can be easily extended to estimate model~\eqref{eq:vc model with t}, which contains time-varying covariates. In fact, we only need to replace the covariates in formulation of the estimator by their time-varying counterparts. The theoretical development is essentially the same. Second, we can allow the varying covariate effect for each component of $\boldsymbol{X}$ to depend on different components of $\boldsymbol{W}$; in practical applications, we may have prior knowledge that the effect of a covariate depends on specific variables. This can be achieved by considering different kernel matrices for each component of $\boldsymbol{X}$. Third, we can consider high-dimensional $\boldsymbol{X}$ and $\boldsymbol{Z}$, which is relevant to omics studies where omics features are often high-dimensional. In practice, it is often the case that only a small subset of features are relevant to a particular outcome. The proposed approaches can be extended to include penalization on the regression parameters/functions, so variable selection can be performed.

\section*{Acknowledgement}

This research was partially supported by the Central Guided Local Science and Technology Development Funds for Research Laboratories (No.2021Szvup145) and the Hong Kong Research Grants Council grant 15303422.

\begin{appendices}
\renewcommand{\thelemma}{\text{C}\arabic{lemma}}

\section{Varying coefficient estimator for discrete $\boldsymbol{W}$}
\label{sec:discrete}

Let $N^{(k)}_{i}(t)$ and $Y^{(k)}_{i}(t)$ denote the event process and at-risk process for the $i$th subject in the $k$th group at time $t$, respectively, where the $k$th group refers to the set of subjects with $\boldsymbol{W}=\boldsymbol{w}_k$. Let $\boldsymbol{X}_i^{(k)}$ be the observed covariate vector for the $i$th subject in the $k$th group ($i=1,\ldots,n_k$) and $\boldsymbol{B}_i=(I(\boldsymbol{W}_i=\boldsymbol{w}_1)\boldsymbol{X}_i^\mathrm{T},\ldots,I(\boldsymbol{W}_i=\boldsymbol{w}_m)\boldsymbol{X}_i^\mathrm{T})^\mathrm{T}$. The average covariate vector at time $t$ is
\begin{align*}
    \overline{\boldsymbol{B}}(t)=\frac{\sum^{n}_{i=1}Y_{i}(t)\boldsymbol{B}_{i}}{\sum^{n}_{i=1}Y_{i}(t)}=&\left(\frac{\sum^{n_{1}}_{i=1}Y^{(1)}_{i}(t)\boldsymbol{X}^{(1)\mathrm{T}}_{i}}{\sum^{n}_{i=1}Y_{i}(t)},\ldots,\frac{\sum^{n_{m}}_{i=1}Y^{(m)}_{i}(t)\boldsymbol{X}^{(m)\mathrm{T}}_{i}}{\sum^{n}_{i=1}Y_{i}(t)}\right)^{\mathrm{T}}\\
    \equiv&\left(\overline{\boldsymbol{X}}^{(1)}(t)^{\mathrm{T}},\ldots,\overline{\boldsymbol{X}}^{(m)}(t)^{\mathrm{T}}\right)^{\mathrm{T}}.
\end{align*}
The Lin and Ying estimator for the regression parameters of the covariate $\boldsymbol{B}$ is $\boldsymbol{V}_{\text{B}}^{-1}\boldsymbol{b}_{\text{B}}$, where $\boldsymbol{b}_{\text{B}}\equiv n^{-1}\sum^{n}_{i=1}\int^{\tau}_{0}\{\boldsymbol{B}_{i}-\overline{\boldsymbol{B}}(t)\}\,\mathrm{d}N_{i}(t)$ is a vector with the $k$th subvector given by 
\begin{align*}
    \frac{1}{n}\sum^{n_{k}}_{i=1}\int^{\tau}_{0}\boldsymbol{X}^{(k)}_{i}\mathrm{d}N^{(k)}_{i}(t)-\frac{1}{n}\sum_{i=1}^n\int^{\tau}_{0}\overline{\boldsymbol{X}}^{(k)}(t)\,\mathrm{d}N_{i}(t),
\end{align*}
and $\boldsymbol{V}_{\text{B}}\equiv n^{-1}\sum^{n}_{i=1}\int^{\tau}_{0}Y_{i}(t)\{\boldsymbol{B}_{i}-\overline{\boldsymbol{B}}(t)\}^{\otimes 2}\,\mathrm{d}t$ is a block matrix with the $(k,\ell)$th block given by
\begin{align*}
    \frac{I(k=\ell)}{n}\int^{\tau}_{0}\sum^{n_{k}}_{i=1}Y^{(k)}_{i}(t)\boldsymbol{X}^{(k)}_{i}\boldsymbol{X}^{(\ell)\mathrm{T}}_{i}\,\mathrm{d}t-\frac{1}{n}\int_0^{\tau}\sum_{i=1}^n Y_i(t)\overline{\boldsymbol{X}}^{(k)}(t)\overline{\boldsymbol{X}}^{(\ell)}(t)^{\mathrm{T}}\,\mathrm{d}t.
\end{align*}

\section{Notations and assumptions}
\label{sec:assumptions}

We impose the following conditions:
\begin{enumerate}[{Condition} B1]
    \item \label{cond:1}
    The marginal density function of $\boldsymbol{W}$, denoted by $f$, is twice continuously differentiable and satisfies $\inf_{\boldsymbol{w}\in[0,1]^q} f(\boldsymbol{w})>0$.
    \item \label{cond:2}
    The conditional density of $(\boldsymbol{X},\boldsymbol{Z})$ given $\boldsymbol{W}=\boldsymbol{w}$ is twice continuously differentiable with respect to $\boldsymbol{w}$. Also, there exists an $M$ such that $P(\Vert\boldsymbol{X}\Vert+\Vert\boldsymbol{Z}\Vert<M)=1$.
    \item \label{cond:3}
    The function $\boldsymbol{\beta}_0(\cdot)$ has continuous second derivative on $[0,1]^q$, and $\int^{\tau}_{0}\lambda_0(t)\,\mathrm{d}t<\infty$.
    \item \label{cond:4}
    The kernel function $K$ is a density function that is symmetric around 0 with bounded derivative, $\int_{\mathbb{R}}x^2K(x)\,\mathrm{d}x<\infty$, and $\int_y^\infty K(x)\,\mathrm{d}x\leq Me^{-ay^2}$ for some $M,a>0$, and large enough $y$.
    \item \label{cond:5}
    As $n\to\infty$, we have $h\to 0$, $m\to\infty$, $h^{q}m\to\infty$, $\log h/(n^{1/2}h^q)\to 0$, and $(nh^q)^{1/2}/(h^qm)\to 0$.
    \item \label{cond:6}
    The $q$-dimension grid $(\boldsymbol{w}_1,\ldots,\boldsymbol{w}_m)$ is the Cartesian product of $q$ grids over the interval $[0,1]$. Each grid consists of a set of points $(a_1,a_2,\ldots,a_{m_k})$ such that $(\max_i \vert a_{i+1}-a_{i}\vert)/(\min_i \vert a_{i+1}-a_{i}\vert)<M$ for all $n$ and some $M<\infty$.
    \item \label{cond:7}
    If $\boldsymbol{\phi}(\boldsymbol{W})^\mathrm{T}\boldsymbol{X}+\boldsymbol{Z}^\mathrm{T}\boldsymbol{a}$ is a constant almost surely for some $\boldsymbol{\phi}\in L^\infty([0,1]^q)$ and real vector $\boldsymbol{a}$, then $\boldsymbol{\phi}(\cdot)=\boldsymbol{0}$ and $\boldsymbol{a}=\boldsymbol{0}$.
    \item \label{cond:8}
    The term $nh^4$ is bounded.
\end{enumerate}

Conditions~B\ref{cond:1}--B\ref{cond:4} are standard conditions in theoretical developments pertaining to kernel methods. Condition~\ref{cond:1} preserves certain smoothness of $f$ so as to apply second order Taylor’s expansion for $f$. Without loss of generality, we consider $\boldsymbol{w}\in[0,1]^q$ in Condition~B\ref{cond:1}. The symmetry of $K$ in Condition~B\ref{cond:4} implies that $\int_{\mathbb{R}}xK(x)\,\mathrm{d}x=0$. Condition~B\ref{cond:7} is for the purpose of model identifiability.

We introduce some notations to be used in the following theoretical development. Let $\boldsymbol{V}=(1,\boldsymbol{X}^\mathrm{T},\boldsymbol{Z}^\mathrm{T})^\mathrm{T}$ and $\boldsymbol{V}_i$ be $\boldsymbol{V}$ for the $i$th subject. For $t\in[0,\tau]$ and $\boldsymbol{w}\in[0,1]^q$, we define
\begin{align*}
    \boldsymbol{S}_n(t,\boldsymbol{w})=&\frac{1}{n}\sum_{i=1}^n K_h(\boldsymbol{W}_i-\boldsymbol{w})Y_i(t)\boldsymbol{V}_i^{\otimes 2}\equiv\left(\begin{matrix} S_{n0}(t,\boldsymbol{w}) & \boldsymbol{S}_{nX}(t,\boldsymbol{w})^\mathrm{T} & \boldsymbol{S}_{nZ}(t,\boldsymbol{w})^\mathrm{T} \\  \boldsymbol{S}_{nX}(t,\boldsymbol{w}) & \boldsymbol{S}_{nX^2}(t,\boldsymbol{w}) & \boldsymbol{S}_{nZX}(t,\boldsymbol{w})^\mathrm{T} \\ \boldsymbol{S}_{nZ}(t,\boldsymbol{w}) & \boldsymbol{S}_{nZX}(t,\boldsymbol{w}) & \boldsymbol{S}_{nZ^2}(t,\boldsymbol{w}) \end{matrix}\right),\\
    \boldsymbol{S}^{*}_{nX^2}(t,\boldsymbol{w})=&\frac{1}{n}\sum^{n}_{i=1}K_h(\boldsymbol{W}_i-\boldsymbol{w})Y_i(t)\lambda(t\mid\boldsymbol{W}_i,\boldsymbol{X}_i,\boldsymbol{Z}_i)\boldsymbol{X}_i^{\otimes 2},
\end{align*}
and their population counterparts
\begin{align*}
    \boldsymbol{s}(t,\boldsymbol{w})=&f(\boldsymbol{w})\mathrm{E}\big(\mathrm{pr}(T\geq t\mid \boldsymbol{W},\boldsymbol{X},\boldsymbol{Z})\boldsymbol{V}^{\otimes 2}\mid \boldsymbol{W}=\boldsymbol{w})\equiv\left(\begin{matrix} s_0(t,\boldsymbol{w}\big) & \boldsymbol{s}_X(t,\boldsymbol{w})^\mathrm{T} & \boldsymbol{s}_Z(t,\boldsymbol{w})^\mathrm{T} \\  \boldsymbol{s}_X(t,\boldsymbol{w}) & \boldsymbol{s}_{X^2}(t,\boldsymbol{w}) & \boldsymbol{s}_{ZX}(t,\boldsymbol{w})^\mathrm{T} \\ \boldsymbol{s}_Z(t,\boldsymbol{w}) & \boldsymbol{s}_{ZX}(t,\boldsymbol{w}) & \boldsymbol{s}_{Z^2}(t,\boldsymbol{w}) \end{matrix}\right),\\
    \boldsymbol{s}^{*}_{X^2}(t,\boldsymbol{w})=&f(\boldsymbol{w})\mathrm{E}\big(\mathrm{pr}(T\geq t\mid \boldsymbol{W},\boldsymbol{X},\boldsymbol{Z})\lambda(t\mid \boldsymbol{W},\boldsymbol{X},\boldsymbol{Z})\boldsymbol{X}^{\otimes 2}\mid \boldsymbol{W}=\boldsymbol{w}\big).
\end{align*}
Similarly, for $t\in[0,\tau]$, we define
\begin{align*}
    S_{n0}(t)=&\frac{1}{n}\sum_{i=1}^n Y_i(t),\quad &s_0(t)=&\mathrm{E}\big(\mathrm{pr}(T\geq t\mid\boldsymbol{W},\boldsymbol{X},\boldsymbol{Z})\big),\\
    \boldsymbol{S}_{nZ}(t)=&\frac{1}{n}\sum_{i=1}^n Y_i(t)\boldsymbol{Z}_i,\quad &\boldsymbol{s}_Z(t)=&\mathrm{E}\big(\mathrm{pr}(T\geq t\mid\boldsymbol{W},\boldsymbol{X},\boldsymbol{Z})\boldsymbol{Z}\big),\\
    \boldsymbol{S}_{nZ^2}(t)=&\frac{1}{n}\sum_{i=1}^n Y_i(t)\boldsymbol{Z}_i^{\otimes 2},\quad &\boldsymbol{s}_{Z^2}(t)=&\mathrm{E}\big(\mathrm{pr}(T\geq t\mid\boldsymbol{W},\boldsymbol{X},\boldsymbol{Z})\boldsymbol{Z}^{\otimes 2}\big).
\end{align*}
Denote
\begin{align*}
    \nu_0=&\int_{\mathbb{R}^q} K(\boldsymbol{z})^2\,\mathrm{d}\boldsymbol{z},\quad\boldsymbol{D}_n(\boldsymbol{w})=\int_0^\tau \boldsymbol{S}_{nX^2}(t,\boldsymbol{w})\,\mathrm{d}t,\quad\boldsymbol{D}(\boldsymbol{w})=\int_0^\tau \boldsymbol{s}_{X^2}(t,\boldsymbol{w})\,\mathrm{d}t,\\
    \boldsymbol{\Psi}(\boldsymbol{w};\boldsymbol{u})=&\int_0^\tau\frac{\boldsymbol{s}_X(t,\boldsymbol{w})\boldsymbol{s}_X(t,\boldsymbol{u})^\mathrm{T}}{s_0(t)}\,\mathrm{d}t+\int_0^\tau\bigg(\boldsymbol{s}_{ZX}(t,\boldsymbol{w})^\mathrm{T}-\frac{\boldsymbol{s}_X(t,\boldsymbol{w})\boldsymbol{s}_Z(t)^\mathrm{T}}{s_0(t)}\bigg)\,\mathrm{d}t\\
    &\times\bigg\{\int_0^\tau\bigg(\boldsymbol{s}_{Z^2}(t)-\frac{\boldsymbol{s}_Z(t)^{\otimes 2}}{s_0(t)}\bigg)\,\mathrm{d}t\bigg\}^{-1}\bigg\{\int_0^\tau\bigg(\boldsymbol{s}_{ZX}(t,\boldsymbol{u})^\mathrm{T}-\frac{\boldsymbol{s}_X(t,\boldsymbol{u})\boldsymbol{s}_Z(t)^\mathrm{T}}{s_0(t)}\bigg)\,\mathrm{d}t\bigg\}^\mathrm{T}.
\end{align*}

\section{Proof of theorems}
\label{sec:proof1}

For any $(a\times b)$ matrices $\boldsymbol{A}_n$ and $\boldsymbol{A}$, we use $\boldsymbol{A}_n=\boldsymbol{A}+O(r_n)$ to denote that $A_{n,ij}=A_{ij}+O(r_n)$ for $i=1,\ldots,a$ and $j=1,\ldots,b$. We present three lemmas that will be used in the proofs of the theorems.
\begin{lemma}
\label{lemma:emp kernel}
Under Conditions~B\ref{cond:1}--B\ref{cond:5},
\[
    \sup_{t\in[0,\tau],\boldsymbol{w}\in[0,1]^q}\left\vert \boldsymbol{S}_n(t,\boldsymbol{w})-\boldsymbol{s}(t,\boldsymbol{w})\right\vert=O_{p}\left(\frac{\log h}{n^{1/2}h^q}\right)+O(h^2),
\]
\[
    \sup_{t\in[0,\tau],\boldsymbol{w}\in[0,1]^q}\left\vert \boldsymbol{S}_{nX^2}^*(t,\boldsymbol{w})-\boldsymbol{s}_{X^2}^*(t,\boldsymbol{w})\right\vert=O_{p}\left(\frac{\log h}{n^{1/2}h^q}\right)+O(h^2).
\]
\end{lemma}

\begin{proof}[Proof of Lemma~C1]
Let $\mathbb{P}$ and $\mathbb{P}_{n}$ denote the true and empirical measures, respectively, and define $\mathbb{G}_{n}=n^{1/2}(\mathbb{P}_{n}-\mathbb{P})$. We can write $\boldsymbol{S}_n(t,\boldsymbol{w})$ as $\mathbb{P}_{n}(K_h(\boldsymbol{W}-\boldsymbol{w})Y(t)\boldsymbol{V}^{\otimes 2})$. By the triangle inequality,
\begin{align*}
    \sup_{t\in[0,\tau],\boldsymbol{w}\in[0,1]^q}\left\vert \boldsymbol{S}_n(t,\boldsymbol{w})-\boldsymbol{s}(t,\boldsymbol{w})\right\vert\leq&\sup_{t\in[0,\tau],\boldsymbol{w}\in[0,1]^q}\left\vert (\mathbb{P}_n-\mathbb{P})\big(K_h(\boldsymbol{W}-\boldsymbol{w})Y(t)\boldsymbol{V}^{\otimes 2}\big)\right\vert\\
    &+\sup_{t\in[0,\tau],\boldsymbol{w}\in[0,1]^q}\left\vert\mathbb{P}\big(K_h(\boldsymbol{W}-\boldsymbol{w})Y(t)\boldsymbol{V}^{\otimes 2}\big)-\boldsymbol{s}(t,\boldsymbol{w})\right\vert.
\end{align*}
Let $\mathcal{F}=\{K_h(\boldsymbol{W}-\boldsymbol{w})Y(t)\boldsymbol{V}^{\otimes 2}:\boldsymbol{w}\in[0,1]^q,t\in[0,\tau]\}$.
We define the uniform entropy integral as
\[
    J(\delta,\mathcal{F},L_2)=\int_0^\delta \bigg(\log\sup_{Q}N(\varepsilon\Vert F\Vert_{Q,2},\mathcal{F},L_2(Q))\bigg)^{1/2}\,\mathrm{d}\varepsilon,
\]
where $N(\varepsilon,\mathcal{F},L_2(Q))$ is the minimal number of $L_2(Q)$-balls of radius $\varepsilon$ needed to cover $\mathcal{F}$, $F$ is an envelop function of $\mathcal{F}$, and $\Vert f\Vert_{Q,r}\equiv(\int\vert f\vert^r\,\mathrm{d}Q)^{1/r}$.

Since $Y(t)$ is stochastically monotone, it follows from Lemma~9.10 of \cite{kosorok2008introduction} that $\mathcal{Y}\equiv\{Y(t):t\in[0,\tau]\}$ is a VC-subgraph class of functions with index 2. By Theorem~2.6.7 of \cite{van1996weak}, the covering number of $\mathcal{Y}$ with $L_2(Q)$ radius less than $\delta$ is of order $O(\delta^{-2})$, and hence $\mathcal{Y}$ has bounded uniform entropy integral. It is direct to verify that for any pairs of $Y(t)$ within the same ball, the $L_2(Q)$ distance of the associated functions in class $\mathcal{F}$ cannot exceed $O(h^{-2q})$, so $N(\delta,\mathcal{F},L_2(Q))\leq O(h^{-2q}\delta^{-2})$. According to Lemma~19.38 of \cite{van2000asymptotic},
\begin{align*}
    \mathrm{E}^{*}_{\mathbb{P}}\Vert\mathbb{G}_{n}\Vert_{\mathcal{F}}&\lesssim J(1,\mathcal{F},L_{2})\Vert F\Vert_{\mathbb{P},2}=\int^{1}_{0}\big(\log O(h^{-2q}\delta^{-2})\big)^{1/2}\,\mathrm{d}\delta h^{-q}=O\left(\frac{\log h}{h^q}\right),
\end{align*}
where $\lesssim$ denotes smaller than up to a constant.
Thus,
\[
    \sup_{t\in[0,\tau],\boldsymbol{w}\in[0,1]^q}\left\vert (\mathbb{P}_n-\mathbb{P})\big(K_h(\boldsymbol{W}-\boldsymbol{w})Y(t)\boldsymbol{V}^{\otimes 2}\big)\right\vert=O_{p}\left(\frac{\log h}{n^{1/2}h^q}\right).
\]
Furthermore,
\[
    \mathbb{P}\big(K_h(\boldsymbol{W}-\boldsymbol{w})Y(t)\boldsymbol{V}^{\otimes 2}\big)=\int_{(\boldsymbol{w}-\mathcal{W})/h} K(\boldsymbol{u})\mathrm{E}\big(Y(t)\boldsymbol{V}^{\otimes 2}\mid \boldsymbol{W}=\boldsymbol{w}+h\boldsymbol{u}\big)f(\boldsymbol{w}+h\boldsymbol{u})\,\mathrm{d}\boldsymbol{u}.
\]
Applying the Taylor series expansion around $\boldsymbol{w}$ (that is, around $h=0$) and taking into account Condition~B4 on $K$, we have
\[
    \mathbb{P}\big(K_h(\boldsymbol{W}-\boldsymbol{w})Y(t)\boldsymbol{V}^{\otimes 2}\big)=f(\boldsymbol{w})\mathrm{E}\big(Y(t)\boldsymbol{V}^{\otimes 2}\mid \boldsymbol{W}=\boldsymbol{w}\big)+O(h^2).
\]
Thus,
\[
    \sup_{t\in[0,\tau],\boldsymbol{w}\in[0,1]^q}\left\vert\mathbb{P}\big(K_h(\boldsymbol{W}-\boldsymbol{w})Y(t)\boldsymbol{V}^{\otimes 2}\big)-\boldsymbol{s}(t,\boldsymbol{w})\right\vert=O(h^2).
\]
Combining the above results, we conclude that
\[
    \sup_{t\in[0,\tau],\boldsymbol{w}\in[0,1]^q}\left\vert \boldsymbol{S}_n(t,\boldsymbol{w})-\boldsymbol{s}(t,\boldsymbol{w})\right\vert=O_{p}\left(\frac{\log h}{n^{1/2}h^q}\right)+O(h^2).
\]
The proofs of the second half of the lemma follows similar arguments and are omitted.
\end{proof}

\begin{lemma}
\label{lemma:emp}
The following results hold:
\[
    \sup_{t\in[0,\tau]}\left\vert \boldsymbol{S}_{nZ^2}(t)-\boldsymbol{s}_{Z^2}(t)\right\vert=O_{p}(n^{-1/2}),
\]
\[
    \sup_{t\in[0,\tau]}\left\vert \boldsymbol{S}_{nZ}(t)-\boldsymbol{s}_Z(t)\right\vert=O_{p}(n^{-1/2}),
\]
\[
    \sup_{t\in[0,\tau]}\left\vert S_{n0}(t)-s_0(t)\right\vert=O_{p}(n^{-1/2}).
\]
\end{lemma}

The proof of Lemma~C2 follows similar arguments as the proof of Lemma~C1 and is omitted.    

\begin{lemma}
\label{lemma:inverse of operator}
    Let $\mathcal{I}$ be the identity operator. For $\boldsymbol{\phi}\in L^2([0,1]^q)$, define the operator $\mathcal{A}$ by
    \[
    \mathcal{A}(\boldsymbol{\phi})(\boldsymbol{w})=\boldsymbol{D}(\boldsymbol{w})^{-1}\int_{\mathcal{W}}\boldsymbol{\Psi}(\boldsymbol{w};\boldsymbol{u})\boldsymbol{\phi}(\boldsymbol{u})\,\mathrm{d}\boldsymbol{u}.
    \]
    Under Condition~B\ref{cond:7}, the inverse operator $(\mathcal{I}-\mathcal{A})^{-1}$ exists.
\end{lemma}

\begin{proof}[Proof of Lemma~C3]
By Theorem~5.2.3 of \cite{debnath2005introduction}, the inverse of the bounded linear operator $(\mathcal{I}-\mathcal{A})$ exists if $(\mathcal{I}-\mathcal{A})(\boldsymbol{\phi})=\boldsymbol{0}$ implies $\boldsymbol{\phi}=\boldsymbol{0}$. Suppose that $(\mathcal{I}-\mathcal{A})(\boldsymbol{\phi})=\boldsymbol{0}$ for some $\boldsymbol{\phi}\in L^2([0,1]^q)$. We have $\boldsymbol{\phi}(\boldsymbol{w})=\mathcal{A}(\boldsymbol{\phi})(\boldsymbol{w})$ for all $\boldsymbol{w}\in[0,1]^q$. Thus,
\begin{align}
\label{eq:inv}
    \int_{\mathcal{W}}\boldsymbol{\phi}(\boldsymbol{w})^\mathrm{T}\boldsymbol{D}(\boldsymbol{w})\boldsymbol{\phi}(\boldsymbol{w})\,\mathrm{d}\boldsymbol{w}-\int_{\mathcal{W}}\int_{\mathcal{W}}\boldsymbol{\phi}(\boldsymbol{w})^\mathrm{T}\boldsymbol{\Psi}(\boldsymbol{w},\boldsymbol{u})\boldsymbol{\phi}(\boldsymbol{u})\,\mathrm{d}\boldsymbol{u}\,\mathrm{d}\boldsymbol{w}=0.
\end{align}
The first term on the left-hand side of \eqref{eq:inv} can be expressed as
\begin{align*}
    &\int_{\mathcal{W}}\boldsymbol{\phi}(\boldsymbol{w})^\mathrm{T}\boldsymbol{D}(\boldsymbol{w})\boldsymbol{\phi}(\boldsymbol{w})\,\mathrm{d}\boldsymbol{w}\\
    =&\int_0^\tau\int_{\mathcal{W}}\boldsymbol{\phi}(\boldsymbol{w})^\mathrm{T}\boldsymbol{s}_{X^2}(t,\boldsymbol{w})\boldsymbol{\phi}(\boldsymbol{w})\,\mathrm{d}\boldsymbol{w}\,\mathrm{d}t\\
    =&\int_0^\tau\int_{\mathcal{W}} f(\boldsymbol{w})\mathrm{E}\big\{\mathrm{pr}(T\geq t\mid \boldsymbol{W},\boldsymbol{X},\boldsymbol{Z})\big(\boldsymbol{\phi}(\boldsymbol{W})^\mathrm{T}\boldsymbol{X}\big)^2\mid \boldsymbol{W}=\boldsymbol{w}\big\}\,\mathrm{d}\boldsymbol{w}\,\mathrm{d}t\\
    =&\int_0^\tau\int_{\mathcal{W}}\int_{\mathcal{X}\times\mathcal{Z}}\widetilde{f}_t(\boldsymbol{w},\boldsymbol{x},\boldsymbol{z})(\boldsymbol{\phi}(\boldsymbol{w})^\mathrm{T}\boldsymbol{x})^2\,\mathrm{d}(\boldsymbol{x},\boldsymbol{z})\,\mathrm{d}\boldsymbol{w}\,\mathrm{d}t\\
    =&\int_0^\tau\widetilde{c}_t\widetilde{\mathrm{E}}_t\big\{\big(\boldsymbol{\phi}(\boldsymbol{W})^\mathrm{T}\boldsymbol{X}\big)^2\big\}\,\mathrm{d}t,
\end{align*}
where $\widetilde{f}_t(\boldsymbol{w},\boldsymbol{x},\boldsymbol{z})=f(\boldsymbol{w},\boldsymbol{x},\boldsymbol{z})\mathrm{pr}(T\geq t\mid \boldsymbol{W}=\boldsymbol{w},\boldsymbol{X}=\boldsymbol{x},\boldsymbol{Z}=\boldsymbol{z})$, $f(\boldsymbol{w},\boldsymbol{x},\boldsymbol{z})$ is the joint density of $(\boldsymbol{W},\boldsymbol{X},\boldsymbol{Z})$, $\widetilde{\mathrm{E}}_t$ is the expectation with respect to $\widetilde{f}_t$, and $\widetilde{c}_t\equiv\{\int_{\mathcal{W}\times\mathcal{X}\times\mathcal{Z}}\widetilde{f}_t(\boldsymbol{w},\boldsymbol{x},\boldsymbol{z})\,\mathrm{d}(\boldsymbol{w},\boldsymbol{x},\boldsymbol{z})\}^{-1}$ is a normalizing constant for $t\in[0,\tau]$.
Following similar arguments, the second term on the left-hand side of \eqref{eq:inv} can be expressed as
\begin{align*}
    &\int_{\mathcal{W}}\int_{\mathcal{W}}\boldsymbol{\phi}(\boldsymbol{w})^\mathrm{T}\boldsymbol{\Psi}(\boldsymbol{w};\boldsymbol{u})\boldsymbol{\phi}(\boldsymbol{u})\,\mathrm{d}\boldsymbol{u}\,\mathrm{d}\boldsymbol{w}\\
    =&\int_0^\tau\widetilde{c}_t\Big\{\widetilde{\mathrm{E}}_t\big(\boldsymbol{\phi}(\boldsymbol{W})^\mathrm{T}\boldsymbol{X}\big)\Big\}^2\,\mathrm{d}t\\
    &+\bigg[\int_0^\tau \widetilde{c}_t\Big\{\widetilde{\mathrm{E}}_t\big(\boldsymbol{\phi}(\boldsymbol{W})^\mathrm{T}\boldsymbol{X}\boldsymbol{Z}^\mathrm{T}\big)-\widetilde{\mathrm{E}}_t\big(\boldsymbol{\phi}(\boldsymbol{W})^\mathrm{T}\boldsymbol{X}\big)\widetilde{\mathrm{E}}_t(\boldsymbol{Z})^\mathrm{T}\Big\}\,\mathrm{d}t\bigg]\\
    &\times\bigg[\int_0^\tau \widetilde{c}_t\Big\{\widetilde{\mathrm{E}}_t(\boldsymbol{Z}^{\otimes 2})-\widetilde{\mathrm{E}}_t(\boldsymbol{Z})^{\otimes 2}\Big\}\,\mathrm{d}t\bigg]^{-1}\\
    &\times\bigg[\int_0^\tau \widetilde{c}_t\Big\{\widetilde{\mathrm{E}}_t\big(\boldsymbol{\phi}(\boldsymbol{W})^\mathrm{T}\boldsymbol{X}\boldsymbol{Z}^\mathrm{T}\big)-\widetilde{\mathrm{E}}_t\big(\boldsymbol{\phi}(\boldsymbol{W})^\mathrm{T}\boldsymbol{X}\big)\widetilde{\mathrm{E}}_t(\boldsymbol{Z})^\mathrm{T}\Big\}\,\mathrm{d}t\bigg]^\mathrm{T}.
\end{align*}
Combining the above results, \eqref{eq:inv} can be expressed as
\begin{align*}
    \int_0^\tau\widetilde{c}_t\widetilde{\mathrm{E}}_t\Bigg\{\bigg(\Big\{\boldsymbol{\phi}(\boldsymbol{W})^\mathrm{T}\boldsymbol{X}-\widetilde{\mathrm{E}}_t\big(\boldsymbol{\phi}(\boldsymbol{W})^\mathrm{T}\boldsymbol{X}\big)\Big\}-\big\{\boldsymbol{Z}-\widetilde{\mathrm{E}}_t(\boldsymbol{Z})\big\}^\mathrm{T}\\
    \times\bigg[\int_0^\tau \widetilde{c}_s\Big\{\widetilde{\mathrm{E}}_s(\boldsymbol{Z}^{\otimes 2})-\widetilde{\mathrm{E}}_s(\boldsymbol{Z})^{\otimes 2}\Big\}\,\mathrm{d}s\bigg]^{-1}&\\
    \times\int_0^\tau\widetilde{c}_s\widetilde{\mathrm{E}}_s\big[\big\{\boldsymbol{Z}-\widetilde{\mathrm{E}}_s(\boldsymbol{Z})\big\}\big\{\boldsymbol{\phi}(\boldsymbol{W})^\mathrm{T}\boldsymbol{X}-\widetilde{\mathrm{E}}_s(\boldsymbol{\phi}(\boldsymbol{W})^\mathrm{T}\boldsymbol{X})\big\}\big]\,\mathrm{d}s\bigg)^2\Bigg\}\,\mathrm{d}t&=0.
\end{align*}
Hence, for $t\in[0,\tau]$,
\begin{align*}
    \widetilde{\mathrm{E}}_t\Bigg\{\bigg(\bigg\{\boldsymbol{\phi}(\boldsymbol{W})^\mathrm{T}\boldsymbol{X}-\widetilde{\mathrm{E}}_t(\boldsymbol{\phi}(\boldsymbol{W})^\mathrm{T}\boldsymbol{X})\bigg\}-\big\{\boldsymbol{Z}-\widetilde{\mathrm{E}}_t(\boldsymbol{Z})\big\}^\mathrm{T}\bigg[\int_0^\tau \widetilde{c}_s\Big\{\widetilde{\mathrm{E}}_s(\boldsymbol{Z}^{\otimes 2})-\widetilde{\mathrm{E}}_s(\boldsymbol{Z})^{\otimes 2}\Big\}\,\mathrm{d}s\bigg]^{-1}&\\
    \times\int_0^\tau\widetilde{c}_s\widetilde{\mathrm{E}}_s\Big[\Big\{\boldsymbol{Z}-\widetilde{\mathrm{E}}_s(\boldsymbol{Z})\Big\}\Big\{\boldsymbol{\phi}(\boldsymbol{W})^\mathrm{T}\boldsymbol{X}-\widetilde{\mathrm{E}}_s\big(\boldsymbol{\phi}(\boldsymbol{W})^\mathrm{T}\boldsymbol{X}\big)\Big\}\Big]\,\mathrm{d}s\bigg)^2\Bigg\}&=0,
\end{align*}
implying that
\begin{align*}
    \Big\{\boldsymbol{\phi}(\boldsymbol{W})^\mathrm{T}\boldsymbol{X}-\widetilde{\mathrm{E}}_t\big(\boldsymbol{\phi}(\boldsymbol{W})^\mathrm{T}\boldsymbol{X}\big)\Big\}-\Big\{\boldsymbol{Z}-\widetilde{\mathrm{E}}_t(\boldsymbol{Z})\Big\}^\mathrm{T}\bigg[\int_0^\tau \widetilde{c}_s\Big\{\widetilde{\mathrm{E}}_s(\boldsymbol{Z}^{\otimes 2})-\widetilde{\mathrm{E}}_s(\boldsymbol{Z})^{\otimes 2}\Big\}\,\mathrm{d}s\bigg]^{-1}&\\
    \times\int_0^\tau\widetilde{c}_s\widetilde{\mathrm{E}}_s\Big[\Big\{\boldsymbol{Z}-\widetilde{\mathrm{E}}_s(\boldsymbol{Z})\Big\}\Big\{\boldsymbol{\phi}(\boldsymbol{W})^\mathrm{T}\boldsymbol{X}-\widetilde{\mathrm{E}}_s\big(\boldsymbol{\phi}(\boldsymbol{W})^\mathrm{T}\boldsymbol{X}\big)\Big\}\Big]\,\mathrm{d}s&=0
\end{align*}
with probability one. Therefore, Condition~B7 implies that $\boldsymbol{\phi}=\boldsymbol{0}$.
\end{proof}

\begin{proof}[Proof of Theorem~\ref{thm:pointest}]

A key step of the proof is to show that for $k=1,\ldots,m$, $\widehat{\boldsymbol{\beta}}(\boldsymbol{w}_k)-\boldsymbol{\beta}_0(\boldsymbol{w}_k)$ satisfies the Fredholm integral equation
\begin{align*}
    (nh^q)^{1/2}\big\{\widehat{\boldsymbol{\beta}}(\boldsymbol{w}_k)-\boldsymbol{\beta}_0(\boldsymbol{w}_k)\big\}=&\boldsymbol{D}(\boldsymbol{w}_k)^{-1}\int_{\mathcal{W}} \boldsymbol{\Psi}(\boldsymbol{w}_k;\boldsymbol{u})(nh^q)^{1/2}\big\{\widehat{\boldsymbol{\beta}}(\boldsymbol{u})-\boldsymbol{\beta}_0(\boldsymbol{u})\big\}\,\mathrm{d}\boldsymbol{u}\\
    &+\boldsymbol{D}(\boldsymbol{w}_k)^{-1}(nh^q)^{1/2}\boldsymbol{M}_n(\boldsymbol{w}_k)\nonumber\\
    &+O_p(n^{1/2}h^{-q/2}m^{-1})+O(n^{1/2}h^{q/2+2})+O_p\left(\frac{\log h}{n^{1/2}h^q}\right)+O(h^2),
\end{align*}
where $\boldsymbol{M}_n(\cdot)$ is defined by \eqref{def:Mn} on page~\pageref{def:Mn}.

Define
\begin{align*}
    \widetilde{\boldsymbol{b}}(\boldsymbol{w}_k)=&\frac{1}{nm}\sum_{i=1}^n\int_0^\tau\big\{K_h(\boldsymbol{W}_i-\boldsymbol{w}_k)\boldsymbol{X}_i-m\widetilde{\boldsymbol{X}}(t,\boldsymbol{w}_k)\big\}\,\mathrm{d}N_i(t),\\
    \widetilde{\boldsymbol{b}}_\alpha=&\frac{1}{n}\sum_{i=1}^n\int_0^\tau\big\{\boldsymbol{Z}_i-\widetilde{\boldsymbol{Z}}(t)\big\}\,\mathrm{d}N_i(t),\\
    \widetilde{\boldsymbol{V}}(\boldsymbol{w}_k,\boldsymbol{w}_\ell)=&
    \frac{I(\boldsymbol{w}_k=\boldsymbol{w}_\ell)}{nm}\sum_{i=1}^n\int_0^{\tau}K_h(\boldsymbol{W}_i-\boldsymbol{w}_k)Y_i(t)\boldsymbol{X}_i^{\otimes 2}\,\mathrm{d}t-\frac{1}{n}\sum_{i=1}^n\int^{\tau}_{0}Y_i(t)\widetilde{\boldsymbol{X}}(t,\boldsymbol{w}_k)\widetilde{\boldsymbol{X}}(t,\boldsymbol{w}_\ell)^\mathrm{T}\,\mathrm{d}t,\\
    \widetilde{\boldsymbol{V}}_{\beta\alpha}(\boldsymbol{w}_k)=&\frac{1}{nm}\sum^{n}_{i=1}\int^{\tau}_{0}K_h(\boldsymbol{W}_i-\boldsymbol{w}_k)Y_{i}(t)\boldsymbol{X}_{i}\big\{\boldsymbol{Z}_{i}-\widetilde{\boldsymbol{Z}}(t)\big\}^{\mathrm{T}}\,\mathrm{d}t,\\
    \widetilde{\boldsymbol{V}}_{\alpha\alpha}=&\frac{1}{n}\sum_{i=1}^n\int_0^\tau Y_i(t)\big\{\boldsymbol{Z}_i-\widetilde{\boldsymbol{Z}}(t)\big\}^{\otimes 2}\,\mathrm{d}t,\\
    \widetilde{\boldsymbol{X}}(t,\boldsymbol{w}_k)=&\frac{\sum_{i=1}^n K_h(\boldsymbol{W}_i-\boldsymbol{w}_k)Y_i(t)\boldsymbol{X}_i}{m\sum_{i=1}^n Y_i(t)}.
\end{align*}
The above terms are obtained by replacing $m^{-1}\sum_{j=1}^m K_h(\boldsymbol{W}_i-\boldsymbol{w}_j)$ in \eqref{eq:estimator} with $1$. Note that
\[
    m^{-1}\sum_{j=1}^m K_h(\boldsymbol{W}_i-\boldsymbol{w}_j)=\int_{\mathcal{W}}K_h(\boldsymbol{W}_i-\boldsymbol{w})\,\mathrm{d}\boldsymbol{w}+O_p(h^{-q}m^{-1})=1+O_p(e^{-ah^{-2}})+O_p(h^{-q}m^{-1})
\]
for $i=1,\ldots,n$. For simplicity, we assume that on the right-hand side above, the second term is dominated by the third term.
Thus, it can be verified that $\boldsymbol{b}(\boldsymbol{w}_k)-\widetilde{\boldsymbol{b}}(\boldsymbol{w}_k)=O_p(h^{-q}m^{-2})$, $\boldsymbol{V}(\boldsymbol{w}_k,\boldsymbol{w}_\ell)-\widetilde{\boldsymbol{V}}(\boldsymbol{w}_k,\boldsymbol{w}_\ell)=O_p(h^{-q}m^{-3})$, $\boldsymbol{V}_{\beta\alpha}(\boldsymbol{w}_k)-\widetilde{\boldsymbol{V}}_{\beta\alpha}(\boldsymbol{w}_k)=O_p(h^{-q}m^{-2})$, $\boldsymbol{V}_{\alpha\alpha}-\widetilde{\boldsymbol{V}}_{\alpha\alpha}=O_p(h^{-q}m^{-1})$, and $\boldsymbol{b}_\alpha-\widetilde{\boldsymbol{b}}_\alpha=O_p(h^{-q}m^{-1})$. 

The proposed estimator is obtained by solving the following system of equations:
\begin{align}
\label{eq:system}
    \left(\begin{matrix} \sum_{j=1}^m \boldsymbol{V}(\boldsymbol{w}_1,\boldsymbol{w}_j)\widehat{\boldsymbol{\beta}}(\boldsymbol{w}_j)+\boldsymbol{V}_{\beta\alpha}(\boldsymbol{w}_1)\widehat{\boldsymbol{\alpha}} \\ \vdots \\ \sum_{j=1}^m \boldsymbol{V}(\boldsymbol{w}_m,\boldsymbol{w}_j)\widehat{\boldsymbol{\beta}}(\boldsymbol{w}_j)+\boldsymbol{V}_{\beta\alpha}(\boldsymbol{w}_m)\widehat{\boldsymbol{\alpha}} \\ \sum_{j=1}^m \boldsymbol{V}_{\beta\alpha}(\boldsymbol{w}_j)^\mathrm{T}\widehat{\boldsymbol{\beta}}(\boldsymbol{w}_j)+\boldsymbol{V}_{\alpha\alpha}\widehat{\boldsymbol{\alpha}} \end{matrix}\right)&=\left(\begin{matrix}\boldsymbol{b}(\boldsymbol{w}_1) \\ \vdots \\ \boldsymbol{b}(\boldsymbol{w}_m) \\ \boldsymbol{b}_\alpha\end{matrix}\right).
\end{align}
For $k=1,\ldots,m$, the $k$th subvector of the left-hand side of \eqref{eq:system} can be expressed as
\begin{align}
\label{eq:LHS}
&\sum_{j=1}^m \boldsymbol{V}(\boldsymbol{w}_k,\boldsymbol{w}_j)\widehat{\boldsymbol{\beta}}(\boldsymbol{w}_j)+\boldsymbol{V}_{\beta\alpha}(\boldsymbol{w}_k)\widehat{\boldsymbol{\alpha}}\nonumber\\
=&\sum_{j=1}^m \widetilde{\boldsymbol{V}}(\boldsymbol{w}_k,\boldsymbol{w}_j)\widehat{\boldsymbol{\beta}}(\boldsymbol{w}_j)+\widetilde{\boldsymbol{V}}_{\beta\alpha}(\boldsymbol{w}_k)\widehat{\boldsymbol{\alpha}}+O_p(h^{-q}m^{-2})\nonumber\\
=&\frac{1}{m}\boldsymbol{D}_n(\boldsymbol{w}_k)\widehat{\boldsymbol{\beta}}(\boldsymbol{w}_k)-\frac{1}{m^2}\sum_{j=1}^m\int_0^\tau\frac{\boldsymbol{S}_{nX}(t,\boldsymbol{w}_k)\boldsymbol{S}_{nX}(t,\boldsymbol{w}_j)}{S_{n0}(t)}^\mathrm{T}\,\mathrm{d}t\widehat{\boldsymbol{\beta}}(\boldsymbol{w}_j)\nonumber\\
&+\frac{1}{m}\int_0^\tau\Big\{\boldsymbol{S}_{nZX}(t,\boldsymbol{w}_k)^\mathrm{T}-\frac{\boldsymbol{S}_{nX}(t,\boldsymbol{w}_k)\boldsymbol{S}_{nZ}(t)}{S_{n0}(t)}^\mathrm{T}\Big\}\,\mathrm{d}t\widehat{\boldsymbol{\alpha}}+O_p(h^{-q}m^{-2}).
\end{align}
By the Doob–Meyer decomposition \cite[Corollary 1.4.1]{fleming2011counting}, the observed event process $N_{i}(t)$ can be uniquely decomposed so that for every $i$, $N_{i}(t)=M_{i}(t)+\int_0^t Y_i(s)\{\lambda_0(s)+\boldsymbol{X}_i^\mathrm{T}\boldsymbol{\beta}_0(\boldsymbol{W}_i)+\boldsymbol{Z}_i^\mathrm{T}\boldsymbol{\alpha}_0\}\,\mathrm{d}s$, where $M_{i}(t)$ is a martingale. Thus, the $k$th subvector of the right-hand side of \eqref{eq:system} can be expressed as
\begin{align}
\label{eq:RHS}
\boldsymbol{b}(\boldsymbol{w}_k)=&\widetilde{\boldsymbol{b}}(\boldsymbol{w}_k)+O_p(h^{-q}m^{-2})\nonumber\\
=&\frac{1}{nm}\sum_{i=1}^n\int_0^\tau\big\{K_h(\boldsymbol{W}_i-\boldsymbol{w}_k)\boldsymbol{X}_i -m\widetilde{\boldsymbol{X}}(t,\boldsymbol{w}_k)\big\}\,\mathrm{d}N_i(t)+O_p(h^{-q}m^{-2})\nonumber\\
=&\frac{1}{nm}\sum_{i=1}^n\int_0^\tau\big\{K_h(\boldsymbol{W}_i-\boldsymbol{w}_k)\boldsymbol{X}_i-m\widetilde{\boldsymbol{X}}(t,\boldsymbol{w}_k)\big\}\,\mathrm{d}M_i(t)\nonumber\\
&+\frac{1}{nm}\sum_{i=1}^n\int_0^\tau K_h(\boldsymbol{W}_i-\boldsymbol{w}_k)Y_i(t)\boldsymbol{X}_i^{\otimes 2}\,\mathrm{d}t\boldsymbol{\beta}_0(\boldsymbol{W}_i)-\frac{1}{m^2}\sum_{j=1}^m\int_0^\tau\frac{\boldsymbol{S}_{nX}(t,\boldsymbol{w}_k)\boldsymbol{S}_{nX}(t,\boldsymbol{w}_j)^\mathrm{T}}{S_{n0}(t)}\,\mathrm{d}t\boldsymbol{\beta}_0(\boldsymbol{w}_j)\nonumber\\
&+\frac{1}{m}\int_0^\tau\Big\{\boldsymbol{S}_{nZX}(t,\boldsymbol{w}_k)^\mathrm{T}-\frac{\boldsymbol{S}_{nX}(t,\boldsymbol{w}_k)\boldsymbol{S}_{nZ}(t)^\mathrm{T}}{S_{n0}(t)}\Big\}\,\mathrm{d}t\boldsymbol{\alpha}_0+O_p(h^{-q}m^{-2})+O(h^2m^{-1}),
\end{align}
because $\boldsymbol{\beta}_0(\boldsymbol{W}_i)= m^{-1}\sum_{j=1}^m K_h(\boldsymbol{W}_i-\boldsymbol{w}_j)\boldsymbol{\beta}_0(\boldsymbol{w}_j)+O_p(h^{-q}m^{-1})+O(h^2)$ for $i=1,\ldots,n$. The last equation in \eqref{eq:system} yields
\begin{align}
\label{eq:alpha hat}
\widehat{\boldsymbol{\alpha}}=&\boldsymbol{V}_{\alpha\alpha}^{-1}\Big\{\boldsymbol{b}_\alpha-\sum_{j=1}^m \boldsymbol{V}_{\beta\alpha}(\boldsymbol{w}_j)^\mathrm{T}\widehat{\boldsymbol{\beta}}(\boldsymbol{w}_j)\Big\}\nonumber\\
=&\widetilde{\boldsymbol{V}}_{\alpha\alpha}^{-1}\Big\{\widetilde{\boldsymbol{b}}_\alpha-\sum_{j=1}^m \widetilde{\boldsymbol{V}}_{\beta\alpha}(\boldsymbol{w}_j)^\mathrm{T}\widehat{\boldsymbol{\beta}}(\boldsymbol{w}_j)\Big\}+O_p(h^{-q}m^{-1})\nonumber\\
=&\bigg[\int_0^\tau\Big\{\boldsymbol{S}_{nZ^2}(t)-\frac{\boldsymbol{S}_{nZ}(t)^{\otimes 2}}{S_{n0}(t)}\Big\}\,\mathrm{d}t\bigg]^{-1}\bigg[\frac{1}{n}\sum_{i=1}^n\int_0^\tau\big\{\boldsymbol{Z}_i-\widetilde{\boldsymbol{Z}}(t)\big\}\,\mathrm{d}N_i(t)\nonumber\\
&-\frac{1}{m}\sum_{j=1}^m\int_0^\tau\Big\{\boldsymbol{S}_{nZX}(t,\boldsymbol{w}_j)^\mathrm{T}-\frac{\boldsymbol{S}_{nX}(t,\boldsymbol{w}_j)\boldsymbol{S}_{nZ}(t)^\mathrm{T}}{S_{n0}(t)}\Big\}^\mathrm{T}\,\mathrm{d}t\widehat{\boldsymbol{\beta}}(\boldsymbol{w}_j)\bigg]+O_p(h^{-q}m^{-1}).
\end{align}
Substituting \eqref{eq:alpha hat} into \eqref{eq:LHS} and setting it to equal \eqref{eq:RHS}, we have
\begin{align}
\label{eq:sample}
&\boldsymbol{D}_n(\boldsymbol{w}_k)(nh^q)^{1/2}\widehat{\boldsymbol{\beta}}(\boldsymbol{w}_k)-\frac{1}{n}\sum_{i=1}^n\int_0^\tau K_h(\boldsymbol{W}_i-\boldsymbol{w}_k)Y_i(t)\boldsymbol{X}_i^{\otimes 2}\,\mathrm{d}t(nh^q)^{1/2}\boldsymbol{\beta}_0(\boldsymbol{W}_i)\nonumber\\
=&(nh^q)^{1/2}\boldsymbol{M}_n(\boldsymbol{w}_k)\nonumber\\
&+\frac{1}{m}\sum_{j=1}^m\Bigg[\int_0^\tau\frac{\boldsymbol{S}_{nX}(t,\boldsymbol{w}_k)\boldsymbol{S}_{nX}(t,\boldsymbol{w}_j)^\mathrm{T}}{S_{n0}(t)}\,\mathrm{d}t\nonumber\\
&+\int_0^\tau\Big\{\boldsymbol{S}_{nZX}(t,\boldsymbol{w}_k)^\mathrm{T}-\boldsymbol{S}_{nX}(t,\boldsymbol{w}_k)\frac{\boldsymbol{S}_{nZ}(t)^\mathrm{T}}{S_{n0}(t)}\Big\}\,\mathrm{d}t\bigg[\int_0^\tau\Big\{\boldsymbol{S}_{nZ^2}(t)-\frac{\boldsymbol{S}_{nZ}(t)^{\otimes 2}}{S_{n0}(t)}\Big\}\,\mathrm{d}t\bigg]^{-1}\nonumber\\
&\times\int_0^\tau\Big\{\boldsymbol{S}_{nZX}(t,\boldsymbol{w}_j)^\mathrm{T}-\boldsymbol{S}_{nX}(t,\boldsymbol{w}_j)\frac{\boldsymbol{S}_{nZ}(t)^\mathrm{T}}{S_{n0}(t)}\Big\}^\mathrm{T}\,\mathrm{d}t\Bigg](nh^q)^{1/2}\big\{\widehat{\boldsymbol{\beta}}(\boldsymbol{w}_j)-\boldsymbol{\beta}_0(\boldsymbol{w}_j)\big\}\nonumber\\
&+O(n^{1/2}h^{q/2+2})+O_p(n^{1/2}h^{-q/2}m^{-1}),
\end{align}
where
\begin{align}
\label{def:Mn}
\boldsymbol{M}_n(\boldsymbol{w})=&\frac{1}{n}\sum_{i=1}^n\int_0^\tau\Bigg(K_h(\boldsymbol{W}_i-\boldsymbol{w})\boldsymbol{X}_i-m\widetilde{\boldsymbol{X}}(t,\boldsymbol{w})\nonumber\\
&-\bigg[\frac{1}{n}\sum_{j=1}^n\int_0^\tau K_h(\boldsymbol{W}_j-\boldsymbol{w})Y_j(t)\boldsymbol{X}_j\big\{\boldsymbol{Z}_j-\widetilde{\boldsymbol{Z}}(t)\big\}^\mathrm{T}\,\mathrm{d}t\bigg]\nonumber\\
                &\times\bigg[\frac{1}{n}\sum_{j=1}^n\int_0^\tau Y_j(t)\big\{\boldsymbol{Z}_j-\widetilde{\boldsymbol{Z}}(t)\big\}^{\otimes 2}\,\mathrm{d}t\bigg]^{-1}\big\{\boldsymbol{Z}_i-\widetilde{\boldsymbol{Z}}(t)\big\}\Bigg)\,\mathrm{d}M_i(t)
                .
\end{align}
The terms in the outer square bracket of \eqref{eq:sample} are the sample counterpart of $\boldsymbol{\Psi}(\boldsymbol{w}_k,\boldsymbol{w}_j)$. By Lemma~\ref{lemma:emp kernel},
\begin{align*}
    &\int_0^\tau\frac{\boldsymbol{S}_{nX}(t,\boldsymbol{w}_k)\boldsymbol{S}_{nX}(t,\boldsymbol{w}_j)^\mathrm{T}}{S_{n0}(t)}\,\mathrm{d}t\nonumber\\
    &+\int_0^\tau\Big\{\boldsymbol{S}_{nZX}(t,\boldsymbol{w}_k)^\mathrm{T}-\boldsymbol{S}_{nX}(t,\boldsymbol{w}_k)\frac{\boldsymbol{S}_{nZ}(t)^\mathrm{T}}{S_{n0}(t)}\Big\}\,\mathrm{d}t\bigg[\int_0^\tau\Big\{\boldsymbol{S}_{nZ^2}(t)-\frac{\boldsymbol{S}_{nZ}(t)^{\otimes 2}}{S_{n0}(t)}\Big\}\,\mathrm{d}t\bigg]^{-1}\nonumber\\
    &\times\int_0^\tau\Big\{\boldsymbol{S}_{nZX}(t,\boldsymbol{w}_j)^\mathrm{T}-\boldsymbol{S}_{nX}(t,\boldsymbol{w}_j)\frac{\boldsymbol{S}_{nZ}(t)^\mathrm{T}}{S_{n0}(t)}\Big\}^\mathrm{T}\,\mathrm{d}t\\
=&\int_0^\tau\frac{\boldsymbol{s}_X(t,\boldsymbol{w}_k)\boldsymbol{s}_X(t,\boldsymbol{w}_j)^\mathrm{T}}{s_0(t)}\,\mathrm{d}t\\
    &+\bigg\{\int_0^\tau\Big(\boldsymbol{s}_{ZX}(t,\boldsymbol{w}_k)^\mathrm{T}-\frac{\boldsymbol{s}_X(t,\boldsymbol{w}_k)\boldsymbol{s}_Z(t)^\mathrm{T}}{s_0(t)}\Big)\,\mathrm{d}t\bigg\}\bigg\{\int_0^\tau\Big(\boldsymbol{s}_{Z^2}(t)-\frac{\boldsymbol{s}_Z(t)^{\otimes 2}}{s_0(t)}\Big)\,\mathrm{d}t\bigg\}^{-1}\\
    &\times\bigg\{\int_0^\tau\Big(\boldsymbol{s}_{ZX}(t,\boldsymbol{w}_j)^\mathrm{T}-\frac{\boldsymbol{s}_X(t,\boldsymbol{w}_j)\boldsymbol{s}_Z(t)^\mathrm{T}}{s_0(t)}\Big)\,\mathrm{d}t\bigg\}^\mathrm{T}+O_p\left(\frac{\log h}{n^{1/2}h^q}\right)+O(h^2)\\
    =&\boldsymbol{\Psi}(\boldsymbol{w}_k,\boldsymbol{w}_j)+O_p\left(\frac{\log h}{n^{1/2}h^q}\right)+O(h^2).
\end{align*}
Also, following similar arguments as in the proof of Lemma~\ref{lemma:emp kernel}, it can be verified that
\begin{align*}
    \frac{1}{n}\sum_{i=1}^n K_h(\boldsymbol{W}_i-\boldsymbol{w}_k)Y_i(t)\boldsymbol{X}_i^{\otimes 2}\boldsymbol{\beta}_0(\boldsymbol{W}_i)=\boldsymbol{s}_{X^2}(t,\boldsymbol{w}_k)\boldsymbol{\beta}_0(\boldsymbol{w}_k)+O_p\left(\frac{\log h}{n^{1/2}h^q}\right)+O(h^2).
\end{align*}
Thus, \eqref{eq:sample} can be expressed as
\begin{align*}
    &\boldsymbol{D}(\boldsymbol{w}_k)(nh^q)^{1/2}\big\{\widehat{\boldsymbol{\beta}}(\boldsymbol{w}_k)-\boldsymbol{\beta}_0(\boldsymbol{w}_k)\big\}\\
    =&(nh^q)^{1/2}\boldsymbol{M}_n(\boldsymbol{w}_k)+\frac{1}{m}\sum_{j=1}^m\boldsymbol{\Psi}(\boldsymbol{w}_k,\boldsymbol{w}_j)(nh^q)^{1/2}\big\{\widehat{\boldsymbol{\beta}}(\boldsymbol{w}_j)-\boldsymbol{\beta}_0(\boldsymbol{w}_j)\big\}\,\mathrm{d}\boldsymbol{w}\\
    &+O_p(n^{1/2}h^{-q/2}m^{-1})+O(n^{1/2}h^{q/2+2})+O_p\left(\frac{\log h}{n^{1/2}h^q}\right)+O(h^2).
\end{align*}
For any function $\boldsymbol{\phi}$ with bounded derivative,
\[
    \frac{1}{m}\sum_{j=1}^m \boldsymbol{\phi}(\boldsymbol{w}_j)^\mathrm{T}(nh^q)^{1/2}\big\{\widehat{\boldsymbol{\beta}}(\boldsymbol{w}_j)-\boldsymbol{\beta}_0(\boldsymbol{w}_j)\big\}=\int_{\mathcal{W}}\boldsymbol{\phi}(\boldsymbol{w})^\mathrm{T}(nh^q)^{1/2}\big\{\widehat{\boldsymbol{\beta}}(\boldsymbol{w})-\boldsymbol{\beta}_0(\boldsymbol{w})\big\}\,\mathrm{d}\boldsymbol{w}+O(m^{-1}).
\]
We can write
\begin{align*}
&(nh^q)^{1/2}\big\{\widehat{\boldsymbol{\beta}}(\boldsymbol{w}_k)-\boldsymbol{\beta}_0(\boldsymbol{w}_k)\big\}\\
=&\,\boldsymbol{D}(\boldsymbol{w}_k)^{-1}(nh^q)^{1/2}\boldsymbol{M}_n(\boldsymbol{w}_k)+\boldsymbol{D}(\boldsymbol{w}_k)^{-1}\int_{\mathcal{W}}\boldsymbol{\Psi}(\boldsymbol{w}_k,\boldsymbol{w})(nh^q)^{1/2}\big\{\widehat{\boldsymbol{\beta}}(\boldsymbol{w})-\boldsymbol{\beta}_0(\boldsymbol{w})\big\}\,\mathrm{d}\boldsymbol{w}\\
&+O_p(n^{1/2}h^{-q/2}m^{-1})+O(n^{1/2}h^{q/2+2})+O_p\left(\frac{\log h}{n^{1/2}h^q}\right)+O(h^2),
\end{align*}
and hence
\begin{align*}
    &(nh^q)^{1/2}(\mathcal{I}-\mathcal{A})\big(\widehat{\boldsymbol{\beta}}-\boldsymbol{\beta}_0\big)(\boldsymbol{w}_k)\\
    =&\,\boldsymbol{D}(\boldsymbol{w}_k)^{-1}(nh^q)^{1/2}\boldsymbol{M}_n(\boldsymbol{w}_k)+O_p(n^{1/2}h^{-q/2}m^{-1})+O(n^{1/2}h^{q/2+2})+O_p\left(\frac{\log h}{n^{1/2}h^q}\right)+O(h^2).
\end{align*}
By Lemma~\ref{lemma:inverse of operator}, $(\mathcal{I}-\mathcal{A})$ is invertible, so
\begin{align*}
    (nh^q)^{1/2}\big(\widehat{\boldsymbol{\beta}}-\boldsymbol{\beta}_0\big)(\boldsymbol{w}_k)=&(nh^q)^{1/2}(\mathcal{I}-\mathcal{A})^{-1}\boldsymbol{D}(\boldsymbol{w}_k)^{-1}\boldsymbol{M}_n(\boldsymbol{w}_k)\\
    &+O_p(n^{1/2}h^{-q/2}m^{-1})+O(n^{1/2}h^{q/2+2})+O_p\left(\frac{\log h}{n^{1/2}h^q}\right)+O(h^2).
\end{align*}
For brevity, we use $(\mathcal{I}-\mathcal{A})^{-1}\boldsymbol{D}(\boldsymbol{w}_k)^{-1}\boldsymbol{M}_n(\boldsymbol{w}_k)$ to represent the operation involving the inverse operator $(\mathcal{I}-\mathcal{A})^{-1}$ applied to the product of functions $\boldsymbol{D}^{-1}$ and $\boldsymbol{M}_n$, followed by evaluation at $\boldsymbol{w}_k$. 
Since $(\mathcal{I}-\mathcal{A})^{-1}=\mathcal{I}+(\mathcal{I}-\mathcal{A})^{-1}\mathcal{A}$, we have
\begin{align}
    (nh^q)^{1/2}\big(\widehat{\boldsymbol{\beta}}-\boldsymbol{\beta}_0\big)(\boldsymbol{w}_k)
    =&\,\boldsymbol{D}(\boldsymbol{w}_k)^{-1}(nh^q)^{1/2}\boldsymbol{M}_n(\boldsymbol{w}_k)+(nh^q)^{1/2}(\mathcal{I}-\mathcal{A})^{-1}\mathcal{A}\boldsymbol{D}(\boldsymbol{w}_k)^{-1}\boldsymbol{M}_n(\boldsymbol{w}_k)\nonumber\\
    &+O_p(n^{1/2}h^{-q/2}m^{-1})+O(n^{1/2}h^{q/2+2})+O_p\left(\frac{\log h}{n^{1/2}h^q}\right)+O(h^2).
    \label{eq:betahat-expand}
\end{align}
By the martingale central limit theorem \citep[Theorem~5.1.1]{fleming2011counting}, for $k=1,\ldots,m$, $(nh^q)^{1/2} \boldsymbol{M}_n(\boldsymbol{w}_k)$ is asymptotically normal with mean zero and variance $\nu_0\int_0^\tau \boldsymbol{s}_{X^2}^*(t,\boldsymbol{w}_k)\,\mathrm{d}t$. By similar arguments, we can show that for $\boldsymbol{w}\in[0,1]^q$, $n^{1/2}\mathcal{A}(\boldsymbol{D}(\boldsymbol{w})^{-1}\boldsymbol{M}_n(\boldsymbol{w}))$ is asymptotically normally distributed. For $\boldsymbol{w},\boldsymbol{w}'\in[0,1]^q$,
\begin{align*}
    &\mathcal{A}(\boldsymbol{D}(\boldsymbol{w})^{-1}\boldsymbol{M}_n(\boldsymbol{w}))-\mathcal{A}(\boldsymbol{D}(\boldsymbol{w}')^{-1}\boldsymbol{M}_n(\boldsymbol{w}'))\\
    =&\int_{\mathcal{W}}\{\boldsymbol{D}(\boldsymbol{w})^{-1}-\boldsymbol{D}(\boldsymbol{w}')^{-1}\}\boldsymbol{\Psi}(\boldsymbol{w};\boldsymbol{u})\boldsymbol{D}(\boldsymbol{u})^{-1}\boldsymbol{M}_n(\boldsymbol{u})\,\mathrm{d}\boldsymbol{u}\\
    &+\int_{\mathcal{W}}\boldsymbol{D}(\boldsymbol{w}')^{-1}\{\boldsymbol{\Psi}(\boldsymbol{w};\boldsymbol{u})-\boldsymbol{\Psi}(\boldsymbol{w}';\boldsymbol{u})\}\boldsymbol{D}(\boldsymbol{u})^{-1}\boldsymbol{M}_n(\boldsymbol{u})\,\mathrm{d}\boldsymbol{u}.
\end{align*}
Since $\boldsymbol{D}(\boldsymbol{w})$ and $\boldsymbol{\Psi}(\boldsymbol{w};\boldsymbol{u})$ are uniformly continuous in $w\in[0,1]^q$, so the process $\{\mathcal{A}(\boldsymbol{D}(\boldsymbol{w})^{-1}\boldsymbol{M}_n(\boldsymbol{w})):\boldsymbol{w}\in[0,1]^q\}$ is uniformly equicontinuous. By Theorem~1.5.4 of \cite{van1996weak}, we conclude that the process $n^{1/2}\mathcal{A}(\boldsymbol{D}(\boldsymbol{w})^{-1}\boldsymbol{M}_n(\boldsymbol{w}))$ converges to a Gaussian process, and thus $(nh^q)^{1/2}(\mathcal{I}-\mathcal{A})^{-1}\mathcal{A}\boldsymbol{D}(\boldsymbol{w}_k)^{-1}\boldsymbol{M}_n(\boldsymbol{w}_k)=O_p(h^{q/2})$. Therefore, all but the first term on the right-hand side of \eqref{eq:betahat-expand} are $o_p(1)$, and
\begin{align*} 
    (nh^q)^{1/2}\big(\widehat{\boldsymbol{\beta}}-\boldsymbol{\beta}_0\big)(\boldsymbol{w}_k)\mathop{\rightarrow}\limits\mathrm{N}(\boldsymbol{0},\boldsymbol{\Sigma}(\boldsymbol{w}_k))
\end{align*}
in distribution, where $\boldsymbol{\Sigma}(\boldsymbol{w}_k)=\boldsymbol{D}(\boldsymbol{w}_k)^{-1}\nu_0\int_0^\tau \boldsymbol{s}_{X^2}^*(t,\boldsymbol{w}_k)\,\mathrm{d}t\boldsymbol{D}(\boldsymbol{w}_k)^{-1}$.
\end{proof}
\end{appendices}

\begin{proof}[Proof of Theorem~2]
Following the proof of Theorem~1, we can write
\begin{align*}
    &n^{1/2}\int_{\mathcal{W}}\boldsymbol{\phi}(\boldsymbol{w})^\mathrm{T}\big\{\widehat{\boldsymbol{\beta}}(\boldsymbol{w})-\boldsymbol{\beta}_0(\boldsymbol{w})\big\}\,\mathrm{d}\boldsymbol{w}\\
    =&n^{1/2}\int_{\mathcal{W}}\boldsymbol{\phi}(\boldsymbol{w})^\mathrm{T}(\mathcal{I}-\mathcal{A})^{-1}\boldsymbol{D}(\boldsymbol{w})^{-1}\boldsymbol{M}_n(\boldsymbol{w})\,\mathrm{d}\boldsymbol{w}\\
    &+O(n^{1/2}h^2)+O_p(n^{1/2}h^{-q}m^{-1})+O_p\left(\frac{\log h}{n^{1/2}h^{3q/2}}\right)+O(h^{2-q/2})\\
    =&n^{1/2}\int_{\mathcal{W}}\boldsymbol{\phi}(\boldsymbol{w})^\mathrm{T}(\mathcal{I}-\mathcal{A})^{-1}\boldsymbol{D}(\boldsymbol{w})^{-1}\\
    &\times\frac{1}{n}\sum_{i=1}^n\int_0^\tau\Bigg(K_h(\boldsymbol{W}_i-\boldsymbol{w})\boldsymbol{X}_i-m\widetilde{\boldsymbol{X}}(t,\boldsymbol{w})\\
    &-\bigg[\frac{1}{n}\sum_{j=1}^n\int_0^\tau K_h(\boldsymbol{W}_j-\boldsymbol{w})Y_j(t)\boldsymbol{X}_j\big\{\boldsymbol{Z}_j-\widetilde{\boldsymbol{Z}}(t)\big\}^\mathrm{T}\,\mathrm{d}t\bigg]\\
    &\times\bigg[\frac{1}{n}\sum_{j=1}^n\int_0^\tau Y_j(t)\big\{\boldsymbol{Z}_j-\widetilde{\boldsymbol{Z}}(t)\big\}^{\otimes 2}\,\mathrm{d}t\bigg]^{-1}\big\{\boldsymbol{Z}_i-\widetilde{\boldsymbol{Z}}(t)\big\}\Bigg)\,\mathrm{d}M_i(t)\,\mathrm{d}\boldsymbol{w}\\
    &+O(n^{1/2}h^2)+O_p(n^{1/2}h^{-q}m^{-1})+O_p\left(\frac{\log h}{n^{1/2}h^{3q/2}}\right)+O(h^{2-q/2})\\
    =&n^{1/2}\frac{1}{n}\sum_{i=1}^n\int_0^\tau\Bigg(\boldsymbol{\phi}(\boldsymbol{W}_i)^\mathrm{T}(\mathcal{I}-\mathcal{A})^{-1}\boldsymbol{D}(\boldsymbol{W}_i)^{-1}\boldsymbol{X}_i\\
    &-\bigg\{\frac{1}{n}\sum_{j=1}^n\boldsymbol{\phi}(\boldsymbol{W}_j)^\mathrm{T}(\mathcal{I}-\mathcal{A})^{-1}\boldsymbol{D}(\boldsymbol{W}_j)^{-1}Y_j(t)\boldsymbol{X}_j\bigg\}\bigg\{\frac{1}{n}\sum_{j=1}^n Y_j(t)\bigg\}^{-1}\\
    &-\bigg[\frac{1}{n}\sum_{j=1}^n\int_0^\tau\boldsymbol{\phi}(\boldsymbol{W}_j)^\mathrm{T}(\mathcal{I}-\mathcal{A})^{-1}\boldsymbol{D}(\boldsymbol{W}_j)^{-1}Y_j(t)\boldsymbol{X}_j\big\{\boldsymbol{Z}_j-\widetilde{\boldsymbol{Z}}(t)\big\}^\mathrm{T}\,\mathrm{d}t\bigg]\\
    &\times\bigg[\frac{1}{n}\sum_{j=1}^n\int_0^\tau Y_j(t)\big\{\boldsymbol{Z}_j-\widetilde{\boldsymbol{Z}}(t)\big\}^{\otimes 2}\,\mathrm{d}t\bigg]^{-1}\big\{\boldsymbol{Z}_i-\widetilde{\boldsymbol{Z}}(t)\big\}\Bigg)\,\mathrm{d}M_i(t)\\
    &+O(n^{1/2}h^2)+O_p(n^{1/2}h^{-q}m^{-1})+O_p\left(\frac{\log h}{n^{1/2}h^{3q/2}}\right)+O(h^{2-q/2}).
\end{align*}
The third equality holds because for $i=1,\ldots,n$, 
\begin{align*}
    \int_{\mathcal{W}}\boldsymbol{\phi}(\boldsymbol{w})^\mathrm{T}(\mathcal{I}-\mathcal{A})^{-1}\boldsymbol{D}(\boldsymbol{w})^{-1}K_h(\boldsymbol{W}_i-\boldsymbol{w})\boldsymbol{X}_i\,\mathrm{d}\boldsymbol{w}=\boldsymbol{\phi}(\boldsymbol{W}_i)^\mathrm{T}(\mathcal{I}-\mathcal{A})^{-1}\boldsymbol{D}(\boldsymbol{W}_i)^{-1}\boldsymbol{X}_i+O_p(h^{-q}m^{-1}).
\end{align*}
By the martingale central limit theorem, $n^{1/2}\int\boldsymbol{\phi}(\boldsymbol{w})^\mathrm{T}\{\widehat{\boldsymbol{\beta}}(\boldsymbol{w})-\boldsymbol{\beta}_0(\boldsymbol{w})\}\,\mathrm{d}\boldsymbol{w}$ is asymptotically normally distributed with mean zero and variance 
\begin{align*}
    \boldsymbol{\Sigma}_\phi=&\int_0^\tau\mathrm{E}\Bigg[\mathrm{pr}(T\geq t\mid\boldsymbol{W},\boldsymbol{X},\boldsymbol{Z})\lambda(t\mid\boldsymbol{W},\boldsymbol{X},\boldsymbol{Z})\bigg(\boldsymbol{\phi}(\boldsymbol{W})^\mathrm{T}(\mathcal{I}-\mathcal{A})^{-1}\boldsymbol{D}(\boldsymbol{W})^{-1}\boldsymbol{X}\\
    &-\mathrm{E}\{\mathrm{pr}(T\geq t\mid\boldsymbol{W},\boldsymbol{X},\boldsymbol{Z})\boldsymbol{\phi}(\boldsymbol{W})^\mathrm{T}(\mathcal{I}-\mathcal{A})^{-1}\boldsymbol{D}(\boldsymbol{W})^{-1}\boldsymbol{X}\}s_0(t)^{-1}\\
    &-\int_0^\tau\mathrm{E}\bigg[\mathrm{pr}(T\geq t\mid\boldsymbol{W},\boldsymbol{X},\boldsymbol{Z})\boldsymbol{\phi}(\boldsymbol{W})^\mathrm{T}(\mathcal{I}-\mathcal{A})^{-1}\boldsymbol{D}(\boldsymbol{W})^{-1}\boldsymbol{X}\Big\{\boldsymbol{Z}-\frac{\boldsymbol{s}_Z(t)}{s_0(t)}\Big\}^\mathrm{T}\bigg]\,\mathrm{d}t\\
    &\times\bigg[\int_0^\tau\Big\{\boldsymbol{s}_{Z^2}(t)-\frac{\boldsymbol{s}_Z(t)^{\otimes 2}}{s_0(t)}\Big\}\,\mathrm{d}t\bigg]^{-1}\Big\{\boldsymbol{Z}-\frac{\boldsymbol{s}_Z(t)}{s_0(t)}\Big\}\bigg)^{\otimes 2}\Bigg]\,\mathrm{d}t.  
    \end{align*}
\end{proof}

\begin{proof}[Proof of Theorem~3]
From the definition of $\widetilde{\boldsymbol{\alpha}}$,
\begin{align}
\label{eq:alpha tilde}
    &\frac{1}{n}\sum_{i=1}^n\int_0^\tau Y_i(t)\big\{\boldsymbol{Z}_i-\widetilde{\boldsymbol{Z}}(t)\big\}^{\otimes 2}\,\mathrm{d}t(\widetilde{\boldsymbol{\alpha}}-\boldsymbol{\alpha}_0)\nonumber\\
    =&\frac{1}{n}\sum_{i=1}^n\int_0^\tau\big\{\boldsymbol{Z}_i-\widetilde{\boldsymbol{Z}}(t)\big\}\,\mathrm{d}N_i(t)-\frac{1}{n}\sum_{i=1}^n\int_0^\tau\big\{\boldsymbol{Z}_i-\widetilde{\boldsymbol{Z}}(t)\big\}Y_i(t)\widehat{\boldsymbol{\beta}}(\boldsymbol{W}_i)^\mathrm{T}\boldsymbol{X}_i\,\mathrm{d}t\nonumber\\
    &-\frac{1}{n}\sum_{i=1}^n\int_0^\tau Y_i(t)\big\{\boldsymbol{Z}_i-\widetilde{\boldsymbol{Z}}(t)\big\}^{\otimes 2}\,\mathrm{d}t\boldsymbol{\alpha}_0\nonumber\\
    =&\frac{1}{n}\sum_{i=1}^n\int_0^\tau\big\{\boldsymbol{Z}_i-\widetilde{\boldsymbol{Z}}(t)\big\}\,\mathrm{d}M_i(t)-\frac{1}{n}\sum_{i=1}^n\int_0^\tau\big\{\boldsymbol{Z}_i-\widetilde{\boldsymbol{Z}}(t)\big\}Y_i(t)\big\{\widehat{\boldsymbol{\beta}}(\boldsymbol{W}_i)-\boldsymbol{\beta}_0(\boldsymbol{W}_i)\big\}^\mathrm{T}\boldsymbol{X}_i\,\mathrm{d}t,
\end{align}
where the second equality is obtained by applying the Doob–Meyer decomposition on $N_i(t)$. Let $\boldsymbol{\phi}(\boldsymbol{w})\equiv(\int_0^\tau f(\boldsymbol{w})\mathrm{E}[\mathrm{pr}(T\geq t\mid\boldsymbol{W},\boldsymbol{X},\boldsymbol{Z})\{\boldsymbol{Z}-\widetilde{\boldsymbol{Z}}(t)\}\boldsymbol{X}^\mathrm{T}\mid\boldsymbol{W}=\boldsymbol{w}]\,\mathrm{d}t)^\mathrm{T}$.
Following the proof of Theorem~2, 
\begin{align*}
    &\frac{1}{n}\sum_{i=1}^n\int_0^\tau\big\{\boldsymbol{Z}_i-\widetilde{\boldsymbol{Z}}(t)\big\}Y_i(t)n^{1/2}\big\{\widehat{\boldsymbol{\beta}}(\boldsymbol{W}_i)-\boldsymbol{\beta}_0(\boldsymbol{W}_i)\big\}^\mathrm{T}\boldsymbol{X}_i\,\mathrm{d}t\\
    =&\int_{\mathcal{W}}\boldsymbol{\phi}(\boldsymbol{w})^\mathrm{T}n^{1/2}\big\{\widehat{\boldsymbol{\beta}}(\boldsymbol{w})-\boldsymbol{\beta}_0(\boldsymbol{w})\big\}\,\mathrm{d}\boldsymbol{w}+O_p(n^{-1/2})\\
    =&n^{1/2}\frac{1}{n}\sum_{i=1}^n\int_0^\tau\Bigg(\boldsymbol{\phi}(\boldsymbol{W}_i)^\mathrm{T}(\mathcal{I}-\mathcal{A})^{-1}\boldsymbol{D}(\boldsymbol{W}_i)^{-1}\boldsymbol{X}_i\\
    &-\bigg\{\frac{1}{n}\sum_{j=1}^n\boldsymbol{\phi}(\boldsymbol{W}_j)^\mathrm{T}(\mathcal{I}-\mathcal{A})^{-1}\boldsymbol{D}(\boldsymbol{W}_j)^{-1}Y_j(t)\boldsymbol{X}_j\bigg\}\bigg\{\frac{1}{n}\sum_{j=1}^n Y_j(t)\bigg\}^{-1}\\
    &-\bigg[\frac{1}{n}\sum_{j=1}^n\int_0^\tau\boldsymbol{\phi}(\boldsymbol{W}_j)^\mathrm{T}(\mathcal{I}-\mathcal{A})^{-1}\boldsymbol{D}(\boldsymbol{W}_j)^{-1}Y_j(t)\boldsymbol{X}_j\big\{\boldsymbol{Z}_j-\widetilde{\boldsymbol{Z}}(t)\big\}^\mathrm{T}\,\mathrm{d}t\bigg]\\
    &\times\bigg[\frac{1}{n}\sum_{j=1}^n\int_0^\tau Y_j(t)\big\{\boldsymbol{Z}_j-\widetilde{\boldsymbol{Z}}(t)\big\}^{\otimes 2}\,\mathrm{d}t\bigg]^{-1}\big\{\boldsymbol{Z}_i-\widetilde{\boldsymbol{Z}}(t)\big\}\Bigg)\,\mathrm{d}M_i(t)\\
    &+O(n^{1/2}h^2)+O_p(n^{1/2}h^{-q}m^{-1})+O_p\left(\frac{\log h}{n^{1/2}h^{3q/2}}\right)+O(h^{2-q/2}),
\end{align*}
which is the sum of orthogonal martingales.Therefore, by the martingale central limit theorem, the right hand side of \eqref{eq:alpha tilde} is asymptotically normally distributed with mean zero and variance
\begin{align*}
    &\int_0^\tau\mathrm{E}\Bigg(\mathrm{pr}(T\geq t\mid\boldsymbol{W},\boldsymbol{X},\boldsymbol{Z})\lambda(t\mid\boldsymbol{W},\boldsymbol{X},\boldsymbol{Z})\bigg[\boldsymbol{Z}-\frac{\boldsymbol{s}_Z(t)}{s_0(t)}-\boldsymbol{\phi}(\boldsymbol{W})^\mathrm{T}(\mathcal{I}-\mathcal{A})^{-1}\boldsymbol{D}(\boldsymbol{W})^{-1}\boldsymbol{X}\\
    &+\mathrm{E}\big\{\mathrm{pr}(T\geq t\mid\boldsymbol{W},\boldsymbol{X},\boldsymbol{Z})\boldsymbol{\phi}(\boldsymbol{W})^\mathrm{T}(\mathcal{I}-\mathcal{A})^{-1}\boldsymbol{D}(\boldsymbol{W})^{-1}\boldsymbol{X}\big\}s_0(t)^{-1}\\
    &+\int_0^\tau\mathrm{E}\bigg(\mathrm{pr}(T\geq t\mid\boldsymbol{W},\boldsymbol{X},\boldsymbol{Z})\boldsymbol{\phi}(\boldsymbol{W})^\mathrm{T}(\mathcal{I}-\mathcal{A})^{-1}\boldsymbol{D}(\boldsymbol{W})^{-1}\boldsymbol{X}\Big\{\boldsymbol{Z}-\frac{\boldsymbol{s}_Z(t)}{s_0(t)}\Big\}^\mathrm{T}\bigg)\,\mathrm{d}t\\
    &\times\bigg[\int_0^\tau\Big\{\boldsymbol{s}_{Z^2}(t)-\frac{\boldsymbol{s}_Z(t)^{\otimes 2}}{s_0(t)}\Big\}\,\mathrm{d}t\bigg]^{-1}\bigg\{\boldsymbol{Z}-\frac{\boldsymbol{s}_Z(t)}{s_0(t)}\bigg\}\bigg]^{\otimes 2}\Bigg)\,\mathrm{d}t.   
\end{align*}
Since
\begin{align*}
    \frac{1}{n}\sum_{i=1}^n\int_0^\tau Y_i(t)\big\{\boldsymbol{Z}_i-\widetilde{\boldsymbol{Z}}(t)\big\}^{\otimes 2}\,\mathrm{d}t=\int_0^\tau\Big\{\boldsymbol{s}_{Z^2}(t)-\frac{\boldsymbol{s}_Z(t)^{\otimes 2}}{s_0(t)}\Big\}\,\mathrm{d}t+O_p(n^{-1/2}),
\end{align*}
the desired result follows.
\end{proof}

\begin{proof}[Proof of Theorem~4]
The estimator for the cumulative baseline hazard $\widehat{\Lambda}(t)$ can be expressed as
\begin{align*}
    &\int_0^t\frac{\sum_{i=1}^n\big\{\mathrm{d}N_i(s)-Y_i(s)\widehat{\boldsymbol{\beta}}(\boldsymbol{W}_i)^\mathrm{T}\boldsymbol{X}_i\,\mathrm{d}s-Y_i(s)\widetilde{\boldsymbol{\alpha}}^\mathrm{T}\boldsymbol{Z}_i\,\mathrm{d}s\big\}}{\sum_{i=1}^n Y_i(s)}\\
    =&\int_0^t\frac{\sum_{i=1}^n \big\{\mathrm{d}M_i(s)+Y_i(s)\,\mathrm{d}\Lambda_0(s\mid \boldsymbol{W}_i,\boldsymbol{X}_i,\boldsymbol{Z}_i)-Y_i(s)\widehat{\boldsymbol{\beta}}(\boldsymbol{W}_i)^\mathrm{T}\boldsymbol{X}_i\,\mathrm{d}s-Y_i(s)\widetilde{\boldsymbol{\alpha}}^\mathrm{T}\boldsymbol{Z}_i\,\mathrm{d}s\big\}}{\sum_{i=1}^n Y_i(s)}\\
    =&\int_0^t\frac{\sum_{i=1}^n \big[\mathrm{d}M_i(s)+Y_i(s)\big\{\mathrm{d}\Lambda_0(s)+(\boldsymbol{\beta}_0(\boldsymbol{W}_i)^\mathrm{T}\boldsymbol{X}_i+\boldsymbol{\alpha}_0^\mathrm{T}\boldsymbol{Z}_i)\mathrm{d}s\big\}\big]}{\sum_{i=1}^n Y_i(s)}\\
    &-\int_0^t\frac{\sum_{i=1}^n \big\{Y_i(s)\widehat{\boldsymbol{\beta}}(\boldsymbol{W}_i)^\mathrm{T}\boldsymbol{X}_i\,\mathrm{d}s+Y_i(s)\widetilde{\boldsymbol{\alpha}}^\mathrm{T}\boldsymbol{Z}_i\,\mathrm{d}s\big\}}{\sum_{i=1}^n Y_i(s)}.
\end{align*}
Thus,
\begin{align*}
    n^{1/2}\big\{\widehat{\Lambda}(t)-\Lambda_0(t)\big\}=&n^{1/2}\int_0^t\frac{\sum_{i=1}^n \mathrm{d}M_i(s)}{\sum_{i=1}^n Y_i(s)}-n^{1/2}\int_0^t\frac{\sum_{i=1}^n \big\{\widehat{\boldsymbol{\beta}}(\boldsymbol{W}_i)-\boldsymbol{\beta}_0(\boldsymbol{W}_i)\big\}^\mathrm{T}\boldsymbol{X}_i Y_i(s)\,\mathrm{d}s}{\sum_{i=1}^n Y_i(s)}\\
    &-\int_0^t\frac{\sum_{i=1}^n Y_i(s)\boldsymbol{Z}_i^\mathrm{T}\,\mathrm{d}s}{\sum_{i=1}^n Y_i(s)}n^{1/2}(\widetilde{\boldsymbol{\alpha}}-\boldsymbol{\alpha}_0).
\end{align*}
Following the proofs of Theorems~2 and 3, the second and third term can be written as sum of orthogonal martingales. Thus, we conclude that $n^{1/2}\{\widehat{\Lambda}(t)-\Lambda_0(t)\}$ converges in distribution to a zero-mean Gaussian process.
\end{proof}

\bibliographystyle{apalike}

\begin{thebibliography}{}

\bibitem[Aalen, 1989]{aalen1989linear}
Aalen, O.~O. (1989).
\newblock A linear regression model for the analysis of life times.
\newblock {\em Statist. Med.}, 8(8):907--925.

\bibitem[Akbani et~al., 2014]{akbani2014pan}
Akbani, R., Ng, P. K.~S., Werner, H.~M., Shahmoradgoli, M., Zhang, F., Ju, Z.,
  Liu, W., Yang, J.-Y., Yoshihara, K., Li, J., et~al. (2014).
\newblock A pan-cancer proteomic perspective on the cancer genome atlas.
\newblock {\em Nat. Commun.}, 5:1--15.

\bibitem[Chen et~al., 2012]{chen2012efficient}
Chen, K., Lin, H., and Zhou, Y. (2012).
\newblock Efficient estimation for the {C}ox model with varying coefficients.
\newblock {\em Biometrika}, 99:379--392.

\bibitem[Cheng et~al., 1995]{cheng1995analysis}
Cheng, S., Wei, L.~J., and Ying, Z. (1995).
\newblock Analysis of transformation models with censored data.
\newblock {\em Biometrika}, 82(4):835--845.

\bibitem[Cox, 1972]{cox1972regression}
Cox, D.~R. (1972).
\newblock {Regression models and life-tables}.
\newblock {\em J. R. Statist. Soc. {\rm B}}, 34(2):187--202.

\bibitem[Debnath and Mikusinski, 2005]{debnath2005introduction}
Debnath, L. and Mikusinski, P. (2005).
\newblock {\em {Introduction to Hilbert Spaces with Applications}}.
\newblock Academic Press.

\bibitem[Fan et~al., 2006]{fan2006local}
Fan, J., Lin, H., and Zhou, Y. (2006).
\newblock {Local partial-likelihood estimation for lifetime data}.
\newblock {\em Ann. Statist.}, 34(1):290--325.

\bibitem[Fan and Zhang, 1999]{fan1999statistical}
Fan, J. and Zhang, W. (1999).
\newblock Statistical estimation in varying coefficient models.
\newblock {\em Ann. Statist.}, 27:1491--1518.

\bibitem[Fleming and Harrington, 2011]{fleming2011counting}
Fleming, T.~R. and Harrington, D.~P. (2011).
\newblock {\em Counting Processes and Survival Analysis}, volume 169.
\newblock John Wiley \& Sons.

\bibitem[Hastie and Tibshirani, 1993]{hastie1993varying}
Hastie, T. and Tibshirani, R. (1993).
\newblock {Varying-coefficient models}.
\newblock {\em J. R. Statist. Soc. {\rm B}}, 55:757--779.

\bibitem[Hunter, 2005]{hunter2005gene}
Hunter, D.~J. (2005).
\newblock Gene--environment interactions in human diseases.
\newblock {\em Nat. Rev. Genet.}, 6:287--298.

\bibitem[Kosorok, 2008]{kosorok2008introduction}
Kosorok, M.~R. (2008).
\newblock {\em {Introduction to Empirical Processes and Semiparametric
  Inference}}.
\newblock Springer.

\bibitem[Landi et~al., 2008]{landi2008gene}
Landi, M.~T., Dracheva, T., Rotunno, M., Figueroa, J.~D., Liu, H., Dasgupta,
  A., Mann, F.~E., Fukuoka, J., Hames, M., Bergen, A.~W., et~al. (2008).
\newblock {Gene expression signature of cigarette smoking and its role in lung
  adenocarcinoma development and survival}.
\newblock {\em PLoS One}, 3(2):e1651.

\bibitem[Li et~al., 2007]{li2007local}
Li, H., Yin, G., and Zhou, Y. (2007).
\newblock Local likelihood with time-varying additive hazards model.
\newblock {\em Can. J. Stat.}, 35(2):321--337.

\bibitem[Lin and Ying, 1994]{lin1994semiparametric}
Lin, D.~Y. and Ying, Z. (1994).
\newblock Semiparametric analysis of the additive risk model.
\newblock {\em Biometrika}, 81(1):61--71.

\bibitem[McKeague and Sasieni, 1994]{mckeague1994partly}
McKeague, I.~W. and Sasieni, P.~D. (1994).
\newblock A partly parametric additive risk model.
\newblock {\em Biometrika}, 81(3):501--514.

\bibitem[Qu et~al., 2018]{qu2018identification}
Qu, L., Song, X., and Sun, L. (2018).
\newblock Identification of local sparsity and variable selection for varying
  coefficient additive hazards models.
\newblock {\em Comput. Statist. Data Anal.}, 125:119--135.

\bibitem[Silverman, 1986]{silverman1986density}
Silverman, B.~W. (1986).
\newblock {\em {Density Estimation for Statistics and Data Analysis}},
  volume~26.
\newblock CRC press.

\bibitem[{The Cancer Genome Atlas Network}, 2013]{tcga2013comprehensive}
{The Cancer Genome Atlas Network} (2013).
\newblock Comprehensive molecular characterization of clear cell renal cell
  carcinoma.
\newblock {\em Nature}, 499(7456):43--49.

\bibitem[Tian et~al., 2005]{tian2005cox}
Tian, L., Zucker, D., and Wei, L. (2005).
\newblock {On the {C}ox model with time-varying regression coefficients}.
\newblock {\em J. Am. Statist. Assoc.}, 100(469):172--183.

\bibitem[van~der Vaart, 2000]{van2000asymptotic}
van~der Vaart, A.~W. (2000).
\newblock {\em {Asymptotic Statistics}}.
\newblock Cambridge University Press.

\bibitem[van~der Vaart and Wellner, 1996]{van1996weak}
van~der Vaart, A.~W. and Wellner, J.~A. (1996).
\newblock {\em {Weak Convergence and Empirical Processes}}.
\newblock Springer.

\bibitem[Yin et~al., 2008]{yin2008partially}
Yin, G., Li, H., and Zeng, D. (2008).
\newblock Partially linear additive hazards regression with varying
  coefficients.
\newblock {\em J. Am. Statist. Assoc.}, 103(483):1200--1213.

\end{thebibliography}

\end{document}